# Perspectives on physics-based one-dimensional modeling of lung physiology

Aranyak Chakravarty[a,b], Debjit Kundu[b], Mahesh V. Panchagnula[b], Alladi Mohan[c], Neelesh A. Patankar[d,1]

[a]School of Nuclear Studies & Application, Jadavpur University, Kolkata, India
[b]Department of Applied Mechanics & Biomedical Engineering, Indian Institute of Technology Madras, Chennai, India
[c]Department of Medicine, Sri Venkateswara Institute of Medical Sciences, Tirupati, India
[d]Department of Mechanical Engineering, Northwestern University, Evanston, IL, USA

## Abstract

The need to understand how infection spreads to the deep lung was acutely realized during the Severe Acute Respiratory Syndrome Coronavirus-2 (SARS-CoV-2) pandemic. The challenge of modeling virus laden aerosol transport and deposition in the airways, coupled with mucus clearance, and infection kinetics, became evident. This perspective provides a consolidated view of coupled one-dimensional physics-based mathematical models to probe multifaceted aspects of lung physiology. Successes of 1D trumpet models in providing mechanistic insights into lung function and optimalities are reviewed while identifying limitations and future directions. Key non-dimensional numbers defining lung function are reported. The need to quantitatively map various pathologies on a physics-based parameter space of non-dimensional numbers (a virtual disease landscape) is noted with an eye on translating modeling to clinical practice. This could aid in disease diagnosis, get mechanistic insights into pathologies, and determine patient specific treatment plan. 1D modeling could be an important tool in developing novel measurement and analysis platforms that could be deployed at point-of-care.

## 1. Introduction

Deep lung infections had occurred commonly during the Severe Acute Respiratory Syndrome Coronavirus2 (SARS-CoV-2) pandemic causing unprecedented number of deaths [1]. One critical question was why this virus was infecting the deep lung much more than other respiratory viruses? How was the virus reaching into the deep lung so efficiently? Was it through the blood or the mucus lining or through aerosol transport in airways or was it growing due to favorable infection kinetics [2–8]? Direct experimental evidence of the underlying mechanism was difficult. The need to rely on indirect evidence together with physical models was evident [5–8]. Lung function is multifaceted and the challenge of modeling virus laden aerosol transport and deposition in the airways, *coupled with* mucus clearance and infection kinetics, became obvious [6]. This perspective reviews and consolidates various mathematical frameworks that can be a powerful tool to achieve this goal in the future.

One dimensional (1D) physics-based models have been providing useful mechanistic insights into lung function and optimalities [8–10]. By 1D models we imply that for each airway the physical variables are functions of the axial coordinate while the radial and circumferential variations are averaged. We provide a unified view of the coupled physics to probe the multifaceted aspect of lung physiology. We interpret the results in a new light. For example, we highlight how steady state solutions during inhalation is a limiting solution for

---

gas invasion into the lung that can provide key insights into lung function. We provide an analytic steady state solution for the first time and provide scaling arguments to guide our intuition about lung physiology. We show results based on known 1D models in literature to highlight the insights based on the effect of various non-dimensional numbers; understanding lung function via non-dimensional numbers is a perspective we emphasize in this article. We discuss how these calculations can be used to gain clinically relevant information for efficient drug delivery via lung or to understand deep lung infection in SARS-CoV-2.

We tabulate the non-dimensional numbers defining lung function and discuss how it could provide a potential pathway to clinical translation. For example, we note an approach to measure and map various pathologies on a physics-based parameter space of non-dimensional numbers. Finally, this article supports the view that 1D modeling could be an important tool in developing novel platforms that can be deployed at point-of-care and that the physics community can contribute toward that vision.

*Background of lung physiology and modeling.* The respiratory system is one of the most exposed organ systems of the human body [11]. It has a complex anatomy which can be broadly subdivided into two regions - the *upper respiratory tract* (URT) and the *lower respiratory tract* (LRT). The URT comprises of the nose, paranasal sinuses, mouth, nasal cavity, pharynx and larynx. The LRT can be further demarcated into the *tracheobronchial tree* and the *alveolar region*. The tracheobronchial tree (also called upper or conducting airways) comprises of the trachea and the dichotomous bifurcating bronchial structure (see Fig. 1a-b). The alveolar region is made of the terminal airways and the alveoli. The bronchial network, along with the alveolar region, is commonly referred to as the *lung*. The alveolar region is sometimes also termed as the *deep lung* due to its terminal location [11, 12]. Pertinent information about lung anatomy is summarised in Table 1.

The large surface area of the airways enables easier respiration through a mechanism which draws in ambient air during inhalation and releases expired air during exhalation. However, it also makes the pulmonary system (and in turn, other internal organs of the body) vulnerable to health hazards from polluting particles and infected droplets, among others, which may enter the respiratory tract during inhalation along with the ambient air. Air, along with suspended particles and droplets, is inhaled into the respiratory tract through the URT, which are transported deeper into the LRT through the conducting airways to reach the deep lung [13]. Gas exchange with the blood stream takes place in the deep lung. The reverse happens during exhalation. While some of the inhaled particles/droplets may get exhaled out, a considerable fraction of these may get deposited in the airway and alveoli as a result of various forces acting on them [14]. Their deposition takes place in the respiratory mucosa - a thin layer of mucosal fluid that separates the airway lumen from the epithelial tissue [6, 15, 16]. The thickness of the mucosal layer is O(50) $\mu$m in the trachea and reduces as one moves deeper becoming negligible in the terminal airways [15]. The mucosal fluid is also subjected to periodic ciliary beating which results in advective transport of the fluid towards the trachea. The deposited particles/droplets are, as such, transported along with the mucosal fluid, in addition to diffusion, resulting in a self-clearance mechanism [15] Pulmonary drug delivery systems also rely on this mechanism to deliver aerosolised medicinal drugs to the lung in a non-invasive manner [17].

A large body of literature exists detailing the extensive investigations that have been carried out on gas exchange dynamics as well as particle/droplet transport and deposition [13, 14, 18, 19]. Nonetheless, certain aspects - such as the dynamics of viral respiratory infections - still require further investigations[20]. Infected virus-laden droplets are the main source for transmission of viral respiratory infections such as SARS-CoV-2 [1, 21–23]. These droplets are formed within the respiratory tract of an infected individual [8, 24, 25] and are subsequently exhaled out during breathing, coughing, sneezing, talking, etc. This leads to formation of turbulent clouds of air containing suspended virus-loaded droplets [26]. These droplets may be inhaled by other individuals causing the inhaled droplets to deposit in the respiratory mucosa releasing the entrained viruses and starting an infection in the new subject [1, 27].

A complete spectrum of respiratory virus transmission includes following components - a) droplet formation from infected respiratory mucosa inside the respiratory tract, b) external transmission of infected droplets through respiratory motions and the effect of preventive measures (such as masks) on such transmission, and c) internal transmission of the inhaled infected droplets within the respiratory tract. A thorough understanding of these components is required in order to fully comprehend the fluid dynamics of respiratory virus transmission.

Extensive studies have been carried out on external transmission of droplets [26, 28], especially since the beginning of the SARS-CoV-2 pandemic. These studies have analysed different physical settings with respect to transmission mechanisms, environmental factors, and physical configurations with a focus on identifying the risk of infection spread and possible preventive measures [29, 30]. While there has been evidence on increased exhaled aerosol during SARS-CoV-2 infection [4], relatively fewer studies have dealt with droplet formation inside the respiratory tract [25, 31–35] which may then be internally transmitted by infected droplets within the respiratory tract [33, 36]. Understanding the mechanism of droplet formation from respiratory mucosa is critical to identifying the conditions which aid droplet formation and hence, increase the risk of infection transmission. Internal transmission of infected droplets within the respiratory tract is a plausible mechanism by which a viral infection may spread to different regions of the respiratory tract, including the alveolar region, where it may cause serious diseases such as pneumonia, acute respiratory distress syndrome etc. [27]. There are not many studies that have taken into account pathogen-specific effects, along with internal droplet transmission, while distinguishing between various types of infections and their spread [37]. This perspective focuses specifically on the fluid dynamics of internal droplet transmission within the respiratory tract in addition to gas exchange. One dimensional (1D) models are surveyed due to their ease of use and efficacy in providing useful clinically relevant mechanistic insights.

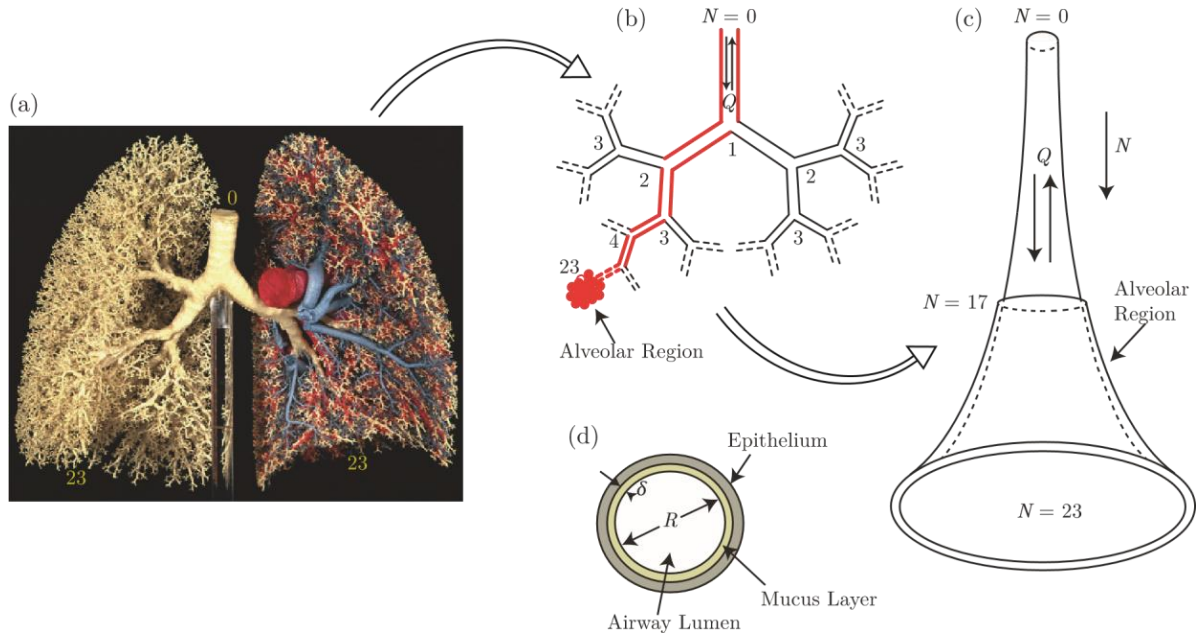

Figure 1: Illustration of the development of the *trumpet* model from an anatomical dichotomous branching network of the lower respiratory tract (LRT). The complete branching network of the LRT is shown in the form of a resin cast in (a) [38] and the corresponding schematic representation is shown in (b). The red-colored pathway highlights a single continuous link of the airway network connecting the extremities. The single continuous link, when lumped considering all branches of each generation, results in a single continuous channel of variable cross-sectional area. This is referred to as the *trumpet* model (c). The respiratory mucus lining the respiratory tract, as considered in the *trumpet* model, is shown in (d). Parts of this figure are reproduced with permission from Hsia et al. [38] and Chakravarty et al. [6, 36].

Table 1: Anatomical parameters for representative lung generations [39].

| $N$ | $N_a$ | $D$ (cm) | $L$ (cm) | $\theta$ (°) | $\phi$ (°) | $\sum S$ (cm$^2$) | $\sum V$ (cm$^3$) |
|---|---|---|---|---|---|---|---|
| 0 | 1 | 2.01 | 10.0 | 0 | 0 | 3.17 | 31.73 |
| 1 | 2 | 1.56 | 4.36 | 33 | 20 | 3.82 | 48.4 |
| 2 | 4 | 1.13 | 1.78 | 34 | 31 | 4.01 | 55.54 |
| 3 | 8 | 0.827 | 0.965 | 22 | 43 | 4.3 | 59.69 |
| 4 | 16 | 0.651 | 0.995 | 20 | 39 | 5.33 | 64.98 |
| 10 | 1024 | 0.198 | 0.556 | 33 | 43 | 31.53 | 148.59 |
| 15 | 32,768 | 0.06 | 0.168 | 51 | 60 | 150.09 | 247.32 |
| 20 | 1,048,576 | 0.044 | 0.07 | 45 | 60 | 1594.39 | 498.36 |
| 23 | 8,388,608 | 0.043 | 0.053 | 45 | 60 | 12,181.95 | 1,692.08 |
| ** | - | - | - | - | - | - | 5563.88 |

$N$: Lung Generation; $N_a$: Number of airways in a generation; $D$: Airway diameter; $L$: Airway length; $\theta$: Branching angle: $\varphi$: Gravity angle; $S$: Total surface area in a generation; $V$: Total volume in a generation; *: Alveoli

---

Different computational techniques have been used in the past to model droplet transport and deposition in the respiratory tract [14, 18, 20, 40]. Geometrical complexity of the respiratory tract, however, makes it difficult (if not impossible) to carry out high resolution computational analyses considering the complete lung geometry. High resolution investigations have, therefore, been carried out separately targeting specific truncated regions of the respiratory tract - the upper bronchial region [41, 42], the central conducting airways [43] or the terminal alveolar region [44–46] - with appropriate boundary conditions. Some studies have also considered combinations of these regions in their analyses by employing different coupling techniques [47–50]. Several studies have also been carried out on droplet transport and deposition in the URT [51]. These studies are useful in modelling geometrical complexities of the respiratory tract and provide useful information regarding the local fluid dynamics as well as insights into the mechanism of droplet deposition in the modelled region. Detailed information of droplet transport and deposition for the complete respiratory tract is expensive and difficult using this approach. Simplified models of the respiratory tract are, therefore, useful when the goal is to capture the key trends of droplet transport and deposition for the complete respiratory tract. Such simplified models are based on morphometry of the respiratory tract [12, 52] and can be broadly classified into semi-empirical regional *compartment* models and one-dimensional *trumpet* models [14]. While these simplified models cannot account for the effects of heterogeneity in the lung, these are extremely useful for understanding the fundamental mechanisms and capturing key trends of aerosol transport and deposition for the whole lung.

The *compartment* models assume that the lung morphology is comprised of four different compartments - extrathoracic, bronchial, bronchiolar, and alveolar regions [14, 20] Separate mathematical models have been developed for each of these compartments, based on analogy with an equivalent electrical circuit, in order to obtain the gas transport and exchange characteristics [53, 54]. Different semi-empirical models are utilised, along with the gas transport models, for determining the deposition of inhaled aerosols within the compartments [14] This technique provides an estimate of the compartment-wise deposition of aerosols.

The *trumpet* model, on the other hand, is a 1D approximation of the anatomical dichotomous branching network of the complete respiratory tract [6, 10, 36, 55–57]. This technique considers the respiratory tract to be a continuous channel of variable cross-sectional area (see Fig. 1c). The length and cross-sectional area of the approximated channel are determined using different empirical correlations based on airway generation number ($N$) and other anatomical parameters. The present work uses a *trumpet* model where the number of branches ($N_l$), length ($L$) and the total cross-sectional area ($A$) at each generation ($N$) is calculated using a power-law function as [6, 36]

$$N_l = 2^N, \; L(N) = L_0 \alpha^N, \; A(N) = A_0 (2\beta)^N, \tag{1}$$

where $L_0$ and $A_0$ are the length and cross-sectional area at $N = 0$ (trachea), respectively. $\alpha$ and $\beta$ are the length-change and area-change factors selected such that the computed length and area at each generation closely matches Weibel's morphometric data [12] (see Fig. 2a). Alveolation of the distal lung airways is considered $N = 17$ onwards, consistent with human lung [12], by considering additional surface area in the relevant generations [6, 36]. The modeled system of airways and alveoli is also assumed to be lined by a thin mucus layer separating the airway lumen from the underlying periciliary layer and the epithelium (see Fig. 1d). The periciliary layer, the epithelium and the ciliary motion driving mucus transport are not explicitly modelled. Instead, mucociliary transport is accounted for by assuming a convective motion in the mucus layer from the deeper generations (larger $N$) towards $N = 0$. The thickness ($\delta$), the total cross-sectional area of the mucus layer ($A_m$), and the convective mucus velocity ($V_m$) at different airway generations are also estimated using power law functions as [6, 36]

$$\delta(N) = \delta_0 \zeta^N, \; A_m(N) = A_{m,0}(2^p \beta \zeta)^N,$$
$$V_m(N) = V_{m,0} \varepsilon^N, \text{ for N < 18,} \tag{2}$$
$$= 0, \text{ for N} \geq 18,$$

where $\delta_0$, $A_{m,0}$, and $V_{m,0}$ are the mucus thickness, area, and velocity at $N = 0$, respectively. The magnitudes of the change factors $\zeta$ and $\varepsilon$ are chosen based on reported data [16, 58, 59]. $V_m$ is assumed to be zero beyond $N = 18$ (Eq 2) due to the absence of appreciable mucociliary transport in the deep lung [15]. $\delta$ and $V_m$ are also assumed to be temporally invariant in this analysis [16].

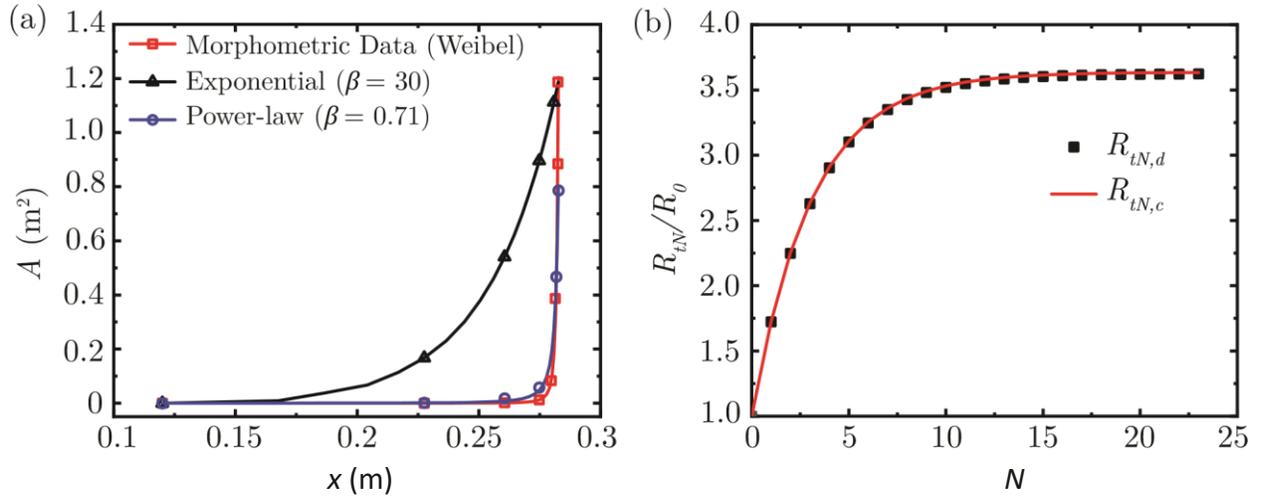

Figure 2: (a) Comparison of area variation in the lung along its length considering the exponential and power-law models with the morphometric data of Weibel [12] (b) Comparison of airway resistance to airflow ($R_{tN}/R_0$) determined using the discrete (Eq. 9) and continuous solutions (Eq. 12) considering $\alpha = 0.73, \beta = 0.71$.

This perspective starts with airway fluid dynamics based on 1D assumptions. It is followed by the transport equations for gas exchange, particle/droplet transport, deposition and clearance and finally, pathogen-specific effects. The goal is to consolidate key insights obtainable from 1D models because of which it has remained a useful model to draw mechanistic insights into lung physiology.

## 2. Airway fluid dynamics

### 2.1. Airflow

In order to develop the airflow transport equations, it is necessary to *a priori* examine the nature of airflow within the lung and make reasonable assumptions. Scaling based on relevant physiological parameters shows that Reynolds number ($Re$) varies between $10^3 - 6 \times 10^{-3}$ within the lung. $Re$ reduces from about $10^3$ at $N = 0$ (trachea) to less than 10 after $N = 13$ (airways) to less than 1 after $N = 17$ (deep lung) [13, 19]. This implies that airflow within most of the lung remains largely laminar. Neglecting turbulence is, thus, a reasonable assumption in most of the lung. Localised air circulation zones can still form near airway bifurcations which can impact particle/aerosol deposition.

The pulsatile nature of breathing introduces unsteadiness to airflow in the upper airways [60]. The corresponding Womersley number ($Wo$) ranges between $6 - 0.1$ with $Wo < 2$ for $N > 2$. This implies that the unsteady pulsatile flow effects exist only in the first couple of generations beyond which the flow remains largely quasi-steady. The transport equations for airflow dynamics are summarized next and the corresponding analysis is presented using the above approximations (also detailed in the subsequent sections).

### 2.1.1. 1D unsteady airflow

The one-dimensional transport equation for airflow in the idealized lung geometry can be expressed as
*Mass transport:*

$$\frac{\partial(\rho A_x)}{\partial t} + \frac{\partial(\rho Q)}{\partial x} = 0 \, ,$$

(3)

*Momentum transport:*

$$\rho \frac{\partial}{\partial t}\left(\frac{Q}{A_x}\right) + \rho \frac{\partial}{\partial x}\left(\frac{1}{2}\left(\frac{Q}{A_x}\right)^2\right) = -\frac{\partial p}{\partial x} - \frac{8\pi\mu\beta_c Q}{N_l A_x^2} \, ,$$

(4)

where $\rho$, $\mu$, $Q$ and $p$ are the air density, air viscosity, volume flow rate of air, and pressure, respectively. $A_x$ is the total cross-sectional area of the airway at a particular position ($x$) defined such that it increases as one goes farther from the trachea (see Fig. 1). $N_l$ is the number of branches in each airway generation (see Eq. 1). The last term on the right hand side of the momentum transport equation (Eq. 4) takes into account the viscous flow resistance and is similar to that obtained in Hagen-Poiseuille flow with a constant cross-sectional area. $\beta_c$ takes into account the unsteady effects of pulsatile flow and can be determined after averaging the Womersley solution. For a parabolic velocity profile $\beta_c = 1$ (Hagen-Poiseuille flow).

### 2.1.2. 1D steady state airflow

As noted earlier, the quasi-steady laminar nature of the air flow through the lung also makes it possible to neglect the inertia terms in Eq. 4 resulting in

$$\frac{\partial p}{\partial x} = -\frac{8\pi\mu\beta_c Q}{N_l A_l^2} \, .$$

(5)

Thus, the airflow in the lung is largely determined by a balance between the pressure gradient and the viscous resistance. This simplification makes it possible to estimate the resistance to airflow in the entire lung. Assuming that the lung has a symmetric branching structure (see Fig. 1), then there are $2^N$ parallel branches at each generation ($N$). It is also assumed that all branches in a single generation are identical. Resistance of a single branch at a particular generation ($N$) can be determined using Eq. 5 as

$$R_{1N} = \frac{\Delta p}{Q} = \frac{8\pi\mu\beta_c L_N}{A_l^2} \, .$$

(6)

Since airflow through the branches is similar to flow through parallel channels [61, 62], the total resistance for generation $N$ can be determined to be

$$R_N = \frac{8\pi\mu\beta_c L_N}{N_l A_l^2} \, .$$

(7)

Using Eq. 1, this can be further reduced to

$$R_N = \left(\frac{8\pi\mu\beta_c L_0}{A_0^2}\right)\left(\frac{\alpha}{2\beta^2}\right)^N = R_0\left(\frac{\alpha}{2\beta^2}\right)^N ,$$

(8)

where, $R_0\left(=\frac{8\pi\mu\beta_c L_0}{A_0^2}\right)$ is the resistance in $N = 0$. The total resistance up to the end of generation $N$ can be determined by adding resistances in series as

$$R_{t N,d} = R_0 + R_0\left(\frac{\alpha}{2\beta^2}\right) + R_0\left(\frac{\alpha}{2\beta^2}\right)^2 + .... + R_0\left(\frac{\alpha}{2\beta^2}\right)^N = R_0\frac{\left[1 - \left(\frac{\alpha}{2\beta^2}\right)^{N+1}\right]}{\left[1 - \left(\frac{\alpha}{2\beta^2}\right)\right]} .$$

(9)

Alternatively, the total airway resistance can also be determined by integrating the resistance per unit length over the entire length of the airway. Using Eq. 1, the total airway length up to the $N^{th}$ generation can be determined to be

$$x_N = L_0 + L_0\alpha + L_0\alpha^2 + ... + L_0\alpha^N = L_0\frac{(1 - \alpha^{N+1})}{(1 - \alpha)} ,$$

(10)

which gives us

$$dx_N = -L_0\frac{\alpha\ln\alpha}{(1 - \alpha)}\alpha^N dN.$$

(11)

Using Eqs. 7 and 11, the total airway resistance up to the end of the generation $N$ can be determined as

$$R_{t N,c} = R_0 + \int_{x_0}^{x_N}\frac{8\pi\mu\beta_c}{N_l A_l^2}dx = R_0 + \int_0^N -R_0\frac{\alpha\ln\alpha}{(1 - \alpha)}\left(\frac{\alpha}{2\beta^2}\right)^N dN = R_0\left[1 + \frac{\alpha\ln\alpha}{(1 - \alpha)\ln\left(\frac{\alpha}{2\beta^2}\right)}\left[1 - \left(\frac{\alpha}{2\beta^2}\right)^N\right]\right]$$

(12)

where $x_0 (= L_0)$ is the length of the $0^{th}$ generation ($N = 0$).

The two solutions obtained by Eqs. 9 and 12 are compared in Fig. 2b for physiologically relevant values of $\alpha$ and $\beta$. The two solutions are in excellent agreement, which allows either solution to be utilised for analysis. It is noted that major deviations of $\beta$ from physiologically relevant values will cause the two solutions to diverge from each other. The technique used in obtaining Eq. 12 will be used to develop and solve other transport equations discussed in this article. It will be found to be useful to get analytic solutions and obtain useful insights into gas transport in the lung.

## 2.2. Gas exchange

Along with airflow dynamics, it is essential to study the mechanism of transport and exchange of various constituent gases (of air) in the lung since breathing induces spatio-temporal variations in gas concentration within the lung [63, 64]. The one-dimensional transport equation for gases in the idealized lung geometry (see Fig. 1) is expressed as

$$\frac{\partial(\rho A_x c_g)}{\partial t} + \frac{\partial(\rho Q c_g)}{\partial x} = \frac{\partial}{\partial x}\left(\rho A_x D_g \frac{\partial c_g}{\partial x}\right) - \rho L_{ex}(c_g - c_\infty) ,$$

(13)

where $c_g$, $Q$, and $D_g$ are the gas concentration (in air), volume flow rate of air during breathing, and gas diffusivity (in air), respectively. $\rho$ is the gas density and $A_x$ is the cross-sectional area of the lung at a particular location $x$. The last term on the right hand side of Eq. 13 takes into account gas exchange with the blood stream (having a gas concentration $c_{g,\infty}$). $L_{ex}$ is defined as the loss coefficient due to gas exchange with the blood stream and is expressed as

$$L_{ex} = 2\pi R D_{ex} ,$$

(14)

where $R$ (= $R_N 2^N$) is the total perimeter in the $N^{th}$ generation and $D_{ex}$ is the rate at which the gas is exchanged with the blood stream. Since gas exchange occurs only in the alveolar region of the lung [11, 64], the loss term is also considered only in the alveolar region of the idealized lung geometry and neglected in the upper airways. Here alveolar capacitance (or lung elasticity) effects could be modeled but are not explicitly elaborated here [20].

Steady and unsteady solutions to Equation 13 are provided in Sections 2.2.1 and 2.2.2, respectively. Fluid motion and hence, gas transport during breathing is inherently unsteady with a time-periodic nature [44] it takes more than 10 breathing cycles (normal breathing cycle ~ 4s) for gas concentration to reach a timeperiodic state. However, the steady-state solution provides useful insights into the transport mechanism, the governing parameters, and it can be used to identify the limiting solution for different flow conditions.

*2.2.1. 1D steady state gas exchange: a reference solution*

The following assumptions are made while solving Eq. 13 using the steady-state approach: a) all parameters are time-invariant, b) airflow within the lung geometry remains uniform and unidirectional, c) gas concentration at the entrance to the trachea ($N = 0$) remains constant, and d) total alveolar gas exchange occurs at the last generation of the lung ($N = 23$).

Considering these assumptions, Eq. 13 simplifies to

$$\frac{\partial}{\partial x}\left(Qc_{g,x} - A_x D_g \frac{\partial c_{g,x}}{\partial x}\right) = 0,$$

(15)

which is solved analytically for two scenarios and numerically for a particular case. The analytic solutions are discussed first, followed by the numerical solution.

*1D analytical solution: exponential variation.* The total cross-sectional area ($A_x$) along the length of the idealised lung geometry may be approximated using an exponential function as

$$A_x = A_0 \exp(\beta x),$$

(16)

where $A_0$ is the cross-sectional area at $x = 0$ and $\beta$ is the area-change factor (see Fig. 2a). The analytical solution of gas concentration obtained considering this approximation is expressed as

$$\lim_{c_{g,\infty},0} \frac{c_{g,x}}{c_{g,0}} = \phi_{g,x} = 1 + \frac{\left[\exp\left[Pe_{g,e}\left(1 - e^{-\beta L}\right)\right] - \exp\left[Pe_{g,e}\left(e^{-\beta x} - e^{-\beta L}\right)\right]\right]\left[1 - \dfrac{Z}{e^{-\beta L}}\right]}{1 - \exp\left[Pe_{g,e}\left(1 - e^{-\beta L}\right)\right]\left[1 - \dfrac{Z}{e^{-\beta L}}\right]},$$

(17)

where $Pe_{g,e}\left(= \dfrac{QL_0}{A_0 D_g}\right)$ and $Z\left(= \dfrac{Q}{A_0 D_{ex}}\right)$ represents the Peclet number for gases (exponential variation) and the gas exchange parameter with the blood stream, respectively. A detailed derivation of the analytical solution and definition of various terms can be found in *Appendix A*.

The magnitude of $Pe_{g,e}$ of interest in respiratory gas transport ranges from 1000 – 64500 based on gas diffusivity in air and air flow rates in breathing. This makes it intractable to determine the magnitude of $\exp\left[Pe_{g,e}\left(1 - e^{-\beta L}\right)\right]$ in Eq. 17. However, it can be used for smaller $Pe_{g,e}$. Additionally, we note that there is a large quantitative difference between the area obtained using the exponential function and the morphometric data (see Fig. 2a; the magnitude at the alveolar end of the lung remains in good agreement). Therefore, the exponential function is not the best fit to the area variation in the lung. However, it models the physics of gas transport in a rapidly diverging duct, which is similar to the lung.

*1D analytical solution: power-law variation.* An alternate approach for a better fit is to approximate the length ($L_N$) and the total cross-sectional area ($A_N$) at each generation ($N$) of the idealised lung geometry using a power-law function (see Section 1: Fig. 2a and Eq. 1). The length-change ($\alpha$) and area-change ($\beta$) factors in Eq. 1 are selected such that the computed length and area at each generation matches Weibel's morphometric data [12] as closely as possible (see Fig. 2a). The airway length ($x$), in terms of the lung generation number ($N$), is given by

$$x_N = \frac{L_0(1 - \alpha^{N+1})}{1 - \alpha}.$$

(18)

Since the length and area variation is assumed to be a function of $N$, the steady-state gas transport equation (Eq. 15) is re-written in terms $N$ as

$$\frac{\partial}{\partial N}\left(Q c_{g,N} - A_N D_g H_N \frac{\partial c_{g,N}}{\partial N}\right) = 0,$$  (19)

using $H_N = \frac{\partial N}{\partial x_N}$. It is to be noted that although $N$ is an integer, it is treated as a continuous variable in all transport equations for computational convenience. The analytic solution for gas concentration following this approach is given by

$$\lim_{c_{g,\infty} \to 0} \frac{c_{g,N}}{c_{g,0}} = \phi_{g,N} = 1 + \frac{\left[\exp\Big(Pe_{g,pl}(\delta^N - 1)\Big) - \exp\Big(Pe_{g,pl}(\delta^M - \delta^N)\Big)\right]\left[1 - \dfrac{Z}{2\beta^M}\right]}{1 - \exp\Big(Pe_{g,pl}(\delta^M - 1)\Big) + Z\dfrac{\exp\Big(Pe_{g,pl}(\delta^M - 1)\Big)}{(2\beta)^M}},$$  (20)

where, $Pe_{g,pl}\left(= \dfrac{QL_0\alpha\ln(\alpha)}{A_0 D_g(1-\alpha)\ln(\alpha/2\beta)}\right)$ represents the Peclet number for gases (power-law variation) and $Z\left(= \dfrac{Q}{A_0 D_{ex}}\right)$ is the gas exchange parameter with the blood stream, respectively. Detailed derivation of the analytical solution and definition of various terms can be found in *Appendix B*. It is to be noted that Eq. 20 also contains terms of the form $\exp\Big(Pe_{g,pl}(\delta^N - 1)\Big)$ (similar to Eq. 17). However, this form is tractable since the small magnitudes of $\delta^N$ negates the large magnitude of $Pe_{g,pl}$ and allows calculation of $c_{g,N}$.

The major governing parameters in Eq. 20 are the Peclet number for gases ($Pe_{g,pl}$) and the gas exchange parameter ($Z$), respectively. Figures 3 and 4 compares the steady-state $\varphi_{g,N}$ for different $Pe_{g,pl}$ ($\sim 1000 - 64500$) and $Z$ ($\sim 500 - 2500$). The range of $Pe_{g,pl}$ considered corresponds to physiological flow rates and gas diffusivity in air. The range of $Z$ is determined by varying the exchange rate ($D_{ex}$).

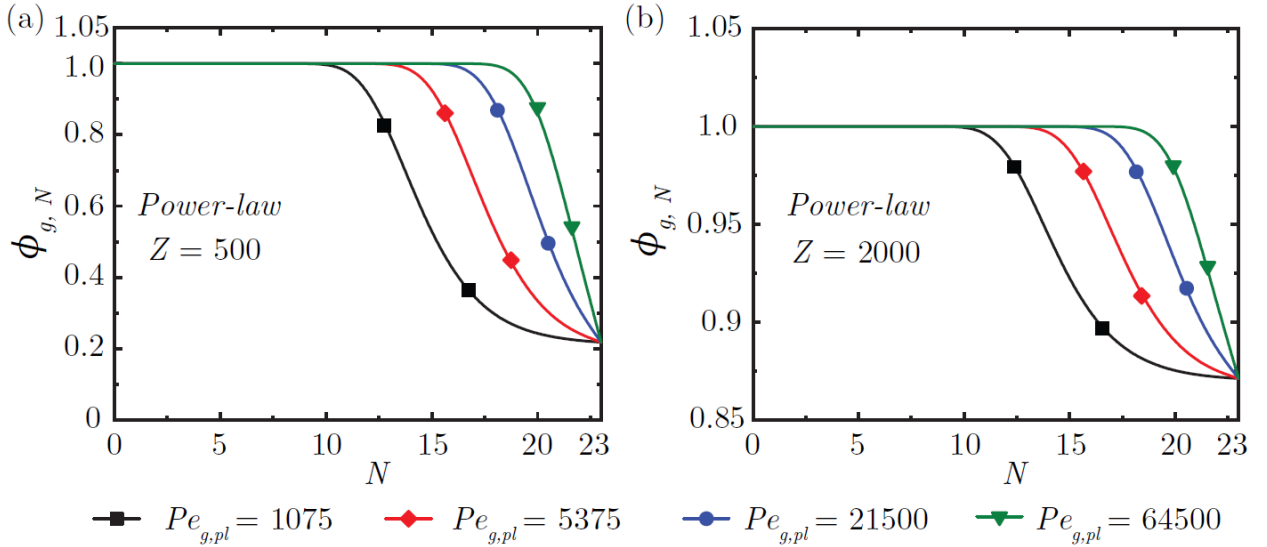

Figure 3: Change in steady-state $\varphi_{g,N}$ within the lung for different $Pe_{g,pl}$ at (a) $Z = 500$ and (b) $Z = 2000$ considering $\alpha = 0.73, \beta = 0.71$. Note the change in scales between the two figures.

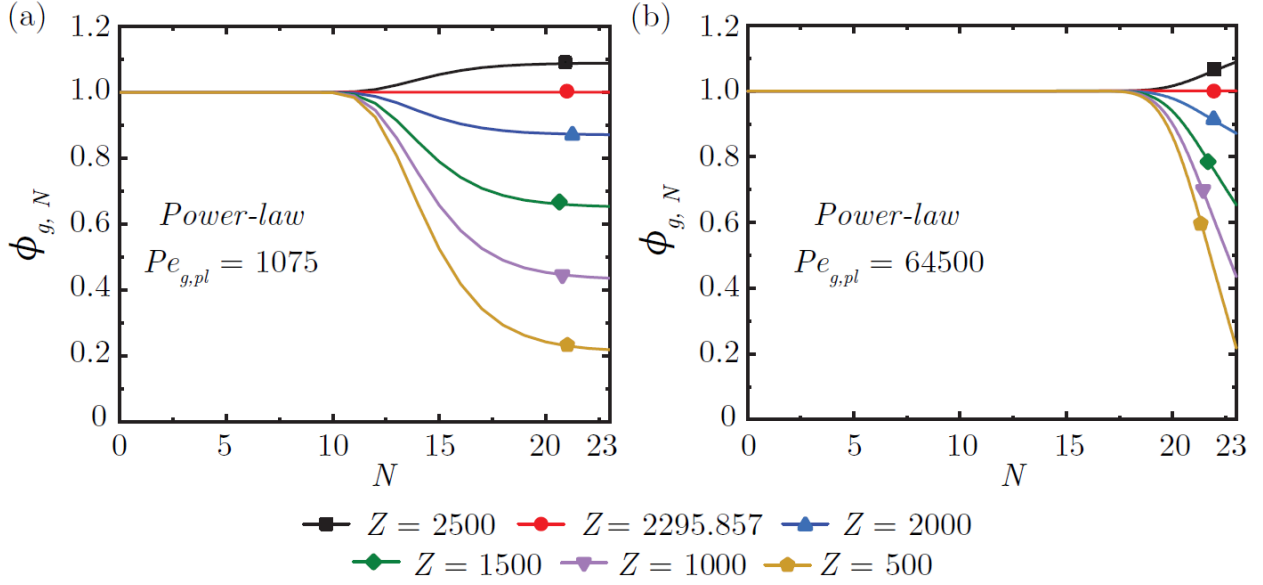

Figure 4: Change in steady-state $\varphi_{g,N}$ within the lung for different $Z$ at (a) $Pe_{g,pl}$ = 1075 and (b) $Pe_{g,pl}$ = 64500 considering $\alpha$ = 0.73, $\beta$ = 0.71.

It is observed from Fig. 3 that, as expected, the gas concentration front propagates deeper into the lung with increase in $Pe_{g,pl}$ for all magnitudes of $Z$. The concentration front demarcates the region where diffusive transport dominates from region where convective transport dominates. For a given gas diffusivity, a larger $Pe_{g,pl}$ results in a greater impact of convection over diffusion and vice-versa. Greater convective transport causes the gas to reach deeper into the lung and hence, the concentration front also propagates towards the deep lung with increase in $Pe_{g,pl}$.

The impact of $Z$ on the gas exchange process is shown in Fig. 4. A larger $Z$ indicates a smaller gas exchange rate and consequently, a greater $\varphi_{g,N}$ at $N$ = 23. A critical magnitude ($Z_{cr}$ = 2295.857) is obtained when no gas exchange takes place with the blood stream and $\varphi_{g,N}$ remains invariant throughout the lung. Any $Z$ larger than this $Z_{cr}$ results in gas accumulation within the lung. It is also evident from Figs. 3 and 4 that the location of the concentration front is determined by the combined effects of $Pe_{g,pl}$ and $Z$. The magnitude of $\varphi_{g,N}$ at $N$ = 23 is, however, independent of $Pe_{g,pl}$ and is solely determined by $Z$.

*1D numerical solution.* The theoretical solutions discussed in the preceding sections considered steadystate transport of gases within the lung for unidirectional air flow. Physiologically, air flow and hence, transport of gases is unsteady and bi-directional with periodic flow reversals depending on the breathing frequency. A numerical method is necessary to obtain an unsteady solution of gas transport equation (Eq. 13). The following scaling parameters (Eq. 21) are used to reduce Eq. 13 to an appropriate dimensionless form (Eq. 22) using the power-law assumption (Eq. 1).

$$\phi_{g,N} = \frac{c_{g,N}}{c_{g,0}}, \tau = \frac{t}{T_b}, Q = Q_{max}q(t), T_a = \frac{L_0 A_0}{|Q_{max}|}, St_a = \frac{T_a}{T_b}, Pe_g = \frac{|Q_{max}|L_0}{A_0 D_g} = Pe_{g,pl}\frac{(1-\alpha)\ln(\alpha/2\beta)}{\alpha\ln(\alpha)q(t)},$$
(21)

where, $Pe_g$ and $St_a$ are the Peclet number for gases and Strouhal number for the airways, respectively. $\varphi_{g,N}$ and $\tau$ denotes the dimensionless gas concentration and time, respectively, while the quantities $T_a$ and $T_b$ represents the convective airflow time-scale and breathing time-scale, respectively. The dimensionless equation used to numerically analyse gas transport within the lung is given by

$$|Pe_g|St_a(2\alpha\beta)^N\frac{\partial\phi_{g,N}}{\partial\tau} = \frac{\partial F_{g,N}}{\partial N} - L'_{ex}\phi_{g,N},$$
(22)

where

$$F_{g,N} = \left[ \left( \left( \frac{2\beta}{\alpha} \right)^N \left( \frac{1-\alpha}{\alpha(\ln\alpha)} \right)^2 \frac{\partial \phi_{g,N}}{\partial N} \right) + \left( |Pe_g| q(t) \left( \frac{1-\alpha}{\alpha \ln(\alpha)} \right) \phi_{g,N} \right) \right] \tag{23},$$

$$L'_{ex} = L_{ex} \frac{L_0^2}{A_0 D_g} \alpha^N = \frac{|Pe_g| \alpha^N}{Z} \left( \frac{2\pi R L_0}{A_0} \right). \tag{24}$$

Equation 22 can be solved numerically [6, 44].

Figure 5 shows the progression of $\phi_{g,N}$ with time advancement within the lung for unidirectional flow. It can be observed that $\phi_{g,N}$ reaches a steady-state within 5 breathing cycles ($\tau \sim 5$) and that the steadystate $\phi_{g,N}$ obtained using the numerical technique matches the theoretical prediction (Eq. 20). Additional comparisons between numerical and theoretical predictions at steady-state are shown in Fig. A1 for the extreme magnitudes of $Pe_{g,pl}$ and $Z$ considered in this study. It can be observed that the numerical solution is able to appropriately capture the trends in $\phi_{g,N}$ for the entire range of $Pe_{g,pl}$ and $Z$ considered in this study for unidirectional flow at steady-state.

### 2.2.2. 1D unsteady gas exchange

Physiologically realistic airflow in the lung is unsteady and bi-directional. For general insights, the numerical model (Eq. 22) can be used to study gas transport within the lung considering physiologically relevant airflow variation during breathing by modeling it as a sinusoidal function i.e. $q(t) = \sin(2\pi t/T_b) = \sin(2\pi\tau)$. Salient results are reproduced in *Appendix C* (Figs. A2-A5) for pertinent parameters.

The wash-in curve (Fig. A2) obtained using the numerical model is similar to that reported in previous studies [65]. It is observed that the wash-in curve reaches a time-periodic state after a few breathing cycles ($\tau \sim 15$) with its frequency corresponding to the breathing cycle, similar to that observed by Chakravarty et al. [44] in a two-dimensional computational study on alveolar gas transport. The nature of the wash-in curve is observed to change in case the gas source becomes unavailable after a certain time (Fig. A3). This is relevant when an individual gets exposed to a certain gas source for a finite period of time ($\tau_{exp}$). While the nature of the curve remains similar as long as the exposure takes place, wash-out of the residual gas starts as soon as the exposure stops. The nature of the wash-out curve is again similar to that observed in literature [63, 66, 67].

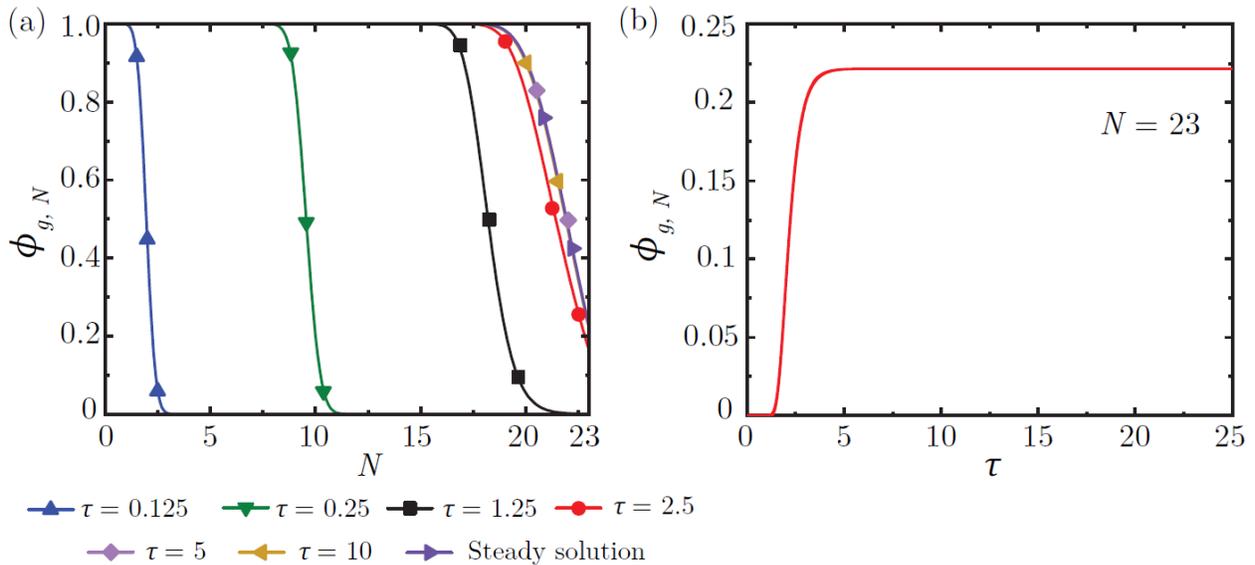

Figure 5: (a) $\phi_{g,N}$ within the lung at different time instances (b) Temporal change in $\phi_{g,N}$ at $N = 23$. The results are shown for $Pe_{g,pl} = 64500$, $Z = 500$, $St_a = 0.01$, $\alpha = 0.73, \beta = 0.71$. It can be observed that the $\phi_{g,N}$ front advances deeper into the lung as time progresses and reaches a steady-state after approximately $\tau = 5$. The steady-state $\phi_{g,N}$ from the numerical solution matches the $\phi_{g,N}$ obtained from the steady-state theoretical solution.

The nature of the wash-in/wash-out curve is also observed to depend on $Pe_g$, $Z$ as well as $St_a$. Other parameters remaining constant, an increase in $Pe_g$ results in a larger advective transport (see Fig. 3) within the lung leading to a relatively greater $\varphi_{g,N}$ at the same time instant and lung generation during inhalation. During exhalation, larger advection leads to faster wash-out and hence, $\varphi_{g,N}$ decreases for a larger $Pe_g$ (Fig. A4a-b). A lower rate of gas exchange (larger $Z$) within the lung results in a larger $\varphi_{g,N}$ at the same time instant and lung generation (Fig. A4c). Washout from the lung, thus, takes a longer time in case of larger $Z$. As expected, the time taken for wash-in increases as $\tau_{exp}$ becomes larger and the washout starts after the corresponding $\tau_{exp}$ ends (Fig. A4d). However, the qualitative nature of the wash-in and washout curves of $\varphi_{g,N}$ remain similar for all $\tau_{exp}$.

The wash-in/wash-out curve is observed to qualitatively change at small $St_a$. A smaller $St_a$ indicates a much larger breathing time-scale ($T_b$) with respect to the advective time-scale. The inhaled gas, as such, is able to progress much deeper into the lung during inhalation as $St_a$ is reduced, other parameters remaining same (see Fig. A5a). Similarly, a smaller $St_a$ allows a longer time for the inhaled gas to get washed out of the lung leading to lower $\varphi_{g,N}$ during exhalation such that minimal wash-out occurs after exposure stops (see Fig. A5b). As $St_a$ increases, wash-out during exposure remains progressively incomplete resulting in some residual $\varphi_{g,N}$ within the lung. The residual $\varphi_{g,N}$ is again transported along with the airflow after exposure ends. As $St_a$ becomes larger, this transport remains restricted to the upper region of the lung and hence, is washed out relatively fast.

## 2.3. Particle transport and deposition

Transport of particles within the lung is fundamentally similar to transport of gases in the lung. Here, the term *particle* is used to denote solid particles as well as liquid droplets/aerosols. The one-dimensional transport equation for particles at any location in the idealised lung geometry can be expressed in a similar manner as that for gases (Eq. 13) as follows [6, 9, 10, 36, 56, 68] -

$$\frac{\partial (Ac_{p,a})}{\partial t} + \frac{\partial (Qc_{p,a})}{\partial x} = \frac{\partial}{\partial x}\left(AD_{p,a}\frac{\partial c_{p,a}}{\partial x}\right) - L_d c_{p,a},\tag{25}$$

where, $c_{p,a}$ and $D_{p,a}$ represents the particle concentration (in airways) and particle diffusivity in air, respectively. $Q$ represents the volume flow rate of air in breathing.

A major physiological difference exists between the gas and particle transport mechanisms. While the gas molecules are exchanged between the blood stream and the lung in the alveoli, inhaled particles/droplets are deposited in the airway mucus throughout the lung. Therefore, the gas exchange term in Eq. 13 is replaced by a deposition term ($L_d c_{p,a}$) in Eq. 25, where $L_d$ is the coefficient of deposition. $L_d$ is calculated taking into account different physical deposition mechanisms viz. diffusional, sedimentation, and impact deposition in the lung airways, as well as diffusional and sedimentation deposition in the alveoli [6, 39].

Equation 25 can be expressed in terms of lung generation number ($N$) assuming a power law variation (see Section 2.2.1) as

$$A_0(2\beta)^N\frac{\partial c_{p,a}}{\partial t} = H\frac{\partial}{\partial N}\left[\left(A_0(2\beta)^N D_{p,a}H\frac{\partial c_{p,a}}{\partial N}\right) - \left(Q_m q(t)c_{p,a}\right)\right] - L_d c_{p,a},\tag{26}$$

where, $q(t)$ represents the temporal sinusoidal function accounting for airflow variation during breathing such that $Q = Q_m q(t)$. Eq. 26 is reduced to its dimensionless form using the following scaling parameters [6]

$$\tau = \frac{t}{T_b}, \phi_{p,a} = \frac{c_{p,a}}{c_{p,a,0}}, T_a = \frac{L_0 A_0}{|Q_m|}, St_a = \frac{T_a}{T_b}, Pe_{p,a} = \frac{|Q_m|L_0}{A_0 D_{p,a}}, D_{p,a} = \frac{k_B T C_S}{3\pi\mu_a d_p},\tag{27}$$

where, $Pe_{p,a}$ and $St_a$ are the Peclet number for particles (in airways) and Strouhal number for the airways, respectively. $\varphi_{p,a}$ and $\tau$ denotes the dimensionless particle concentration (in airways) and time, respectively, while the quantities $T_a$ and $T_b$ represents the convective airflow time-scale and breathing time-scale, respectively. The expression of $D_{p,a}$ is based on the Stokes-Einstein relation [44], where $C_s$ represents the Cunningham slip correction, $T$ represents the ambient temperature, $\mu_a$ denotes air viscosity, and $d_p$ denotes the particle diameter.

The dimensionless equation is expressed as [6, 36]

$$|Pe_{p,a}|St_a(2\alpha\beta)^N\frac{\partial(\phi_{p,a})}{\partial\tau} = \frac{\partial F_{p,a}}{\partial N} - L'_d\phi_{p,a},$$

(28)

where, $L^0_d$ represents the dimensionless form of droplet deposition coefficient ($L_d$) and $F_{p,a}$ represents the total particle flux in airways. These are expressed as follows -

$$L'_d = L_d\frac{L_0^2}{A_0 D_{p,a}}\alpha^N,$$

(29)

$$F_{p,a} = \left[\left(\left(\frac{2\beta}{\alpha}\right)^N\left(\frac{1-\alpha}{\alpha(\ln\alpha)}\right)^2\frac{\partial\phi_{p,a}}{\partial N}\right) + \left(|Pe_{p,a}|q(t)\left(\frac{1-\alpha}{\alpha\,\ln(\alpha)}\right)\phi_{p,a}\right)\right]$$

(30)

The transport equations (Eq. 25 and 28) are formulated based on the assumption that the particles are monodispersed, do not undergo coagulation, and are decoupled from airflow in the lung. It is also assumed that there is no additional source of particles present within the lung and the particles are either deposited in the airway mucus or washed out of the airways.

The major mechanisms of particle deposition in the lung have been identified in the literature as diffusion, sedimentation and impaction of the particles in the airways, as well as diffusion and sedimentation of the particles in the alveoli [14, 57]. Different empirical models available in the literature have been used to estimate $L'_d$ by suitably modifying them to maintain consistency with the present mathematical formulation (see Chakravarty et al. [6] for details). The final dimensionless form of these deposition models are expressed as follows [6]-

*Diffusional deposition in the airways*

$$L'_{d,D} = L_{d,D}\frac{L_0^2}{A_0 D_{p,a}}\alpha^N = \left(\frac{L_0}{R_0}\right)^2(2\alpha)^N(3.66 + 22.305 + 57)$$

(31)

*Sedimentation deposition in the airways*

$$L'_{d,s} = L_{d,s}\frac{L_0^2}{A_0 D_{p,a}}\alpha^N = \frac{1}{3}\left(\frac{L_0}{R_0}\right)^2(2\alpha\sqrt{\beta})^N S_g\cos(\psi_N),$$

(32)

where, $S_g$ is defined as the sedimentation parameter and expressed as

$$S_g = \frac{R_0\rho_p d_p^3 g}{k_B T}.$$

(33)

*Impact deposition in the airways*

$$L'_{d,i} = L_{d,i}\frac{L_0^2}{A_0 D_{p,a}}\alpha^N = |Pe_{p,a}|q(t)\ln(f_i^N(\theta, St))\frac{(1-\alpha)}{\alpha\,\ln(\alpha)}$$

(34)

*Diffusional deposition in the alveoli*

$$L'_{d,D,alv} = \gamma_N\eta_{D,alv}|Pe_{p,a}|q(t)\left(\frac{1-\alpha}{-\alpha\ln(\alpha)}\right),$$

(35)

where, $\gamma_N$ denotes the fraction of alveolated area in the corresponding generation [6] and $\eta_{D,alv}$ denotes the diffusional deposition efficiency in the alveoli. $\eta_{D,alv}$ is expressed as [57]

$$\eta_{D,alv} = 1 - \frac{6}{\pi^2}\sum\frac{1}{k^2}exp\left[-\frac{4k^2 t D_p}{d_{eq}^2}\right]$$

(36)

*Sedimentation deposition in the alveoli*

$$L'_{d,s,alv} = \gamma_N \eta_{s,alv} |Pe_{p,a}| q(t) \left( \frac{1 - \alpha}{-\alpha \ln(\alpha)} \right),$$

(37)

where, $\gamma_N$ and $\eta_{s,alv}$ denotes the fraction of alveolated area in the corresponding generation [6] and sedimentation deposition efficiency in the alveoli, respectively. $\eta_{s,alv}$ is expressed as [57]

$$\eta_{s,alv} = \left[ 1 + \min\left( \frac{d_s}{d_{eq}}, 1 \right) \right]^2 \left[ 1 - 0.5\min\left( \frac{d_s}{d_{eq}}, 1 \right) \right]^2 - 1.$$

(38)

Figure 6a illustrates the transport of particles within the lung airways in a particular breathing cycle. Once inhaled, the particles are carried along with the airflow inwards into the lung during the inhalation phase of the breathing cycle. The particle concentration ($\varphi_{p,a}$) reaches its deepest region in the lung at the end of inhalation. This depends on $Pe_{p,a}$ which is determined by the airflow velocity and particle diffusivity. During the exhalation phase of the breathing cycle, the particles are transported towards the upper regions of the lung and are ultimately washed out of the lung. However, it can be observed from Fig. 6a that all inhaled particles may not get washed out in a single breathing cycle leading to some residual $\varphi_{p,a}$ within the lung at the end of exhalation. These residual particles are again transported inwards into the lung along with any additional inhaled particles [6, 44].

As discussed, the inhaled particles may also get deposited in the airway mucus while being transported within the lung. As a result of deposition, the net $\varphi_{p,a}$ decreases which is apparent from the comparative plot in Fig. 6a. The total amount of particle deposited in the airway mucus is shown in Fig. 6b for a representative case. It can be observed that the deposition is non-uniform and it increases towards the deep lung reaching a maximum at $N = 22$. This trend of deposition is, however, observed to be dependent on $Pe_{p,a}$, $St_a$ and the exposure duration [6]. Exposure duration ($\tau_{exp}$) is observed to only quantitatively influence particle deposition (other parameters remaining same) with the total deposition increasing linearly as $\tau_{exp}$ becomes longer (see Fig. A7b). However, $Pe_{p,a}$ as well as $St_a$ are observed to qualitatively influence particle deposition.

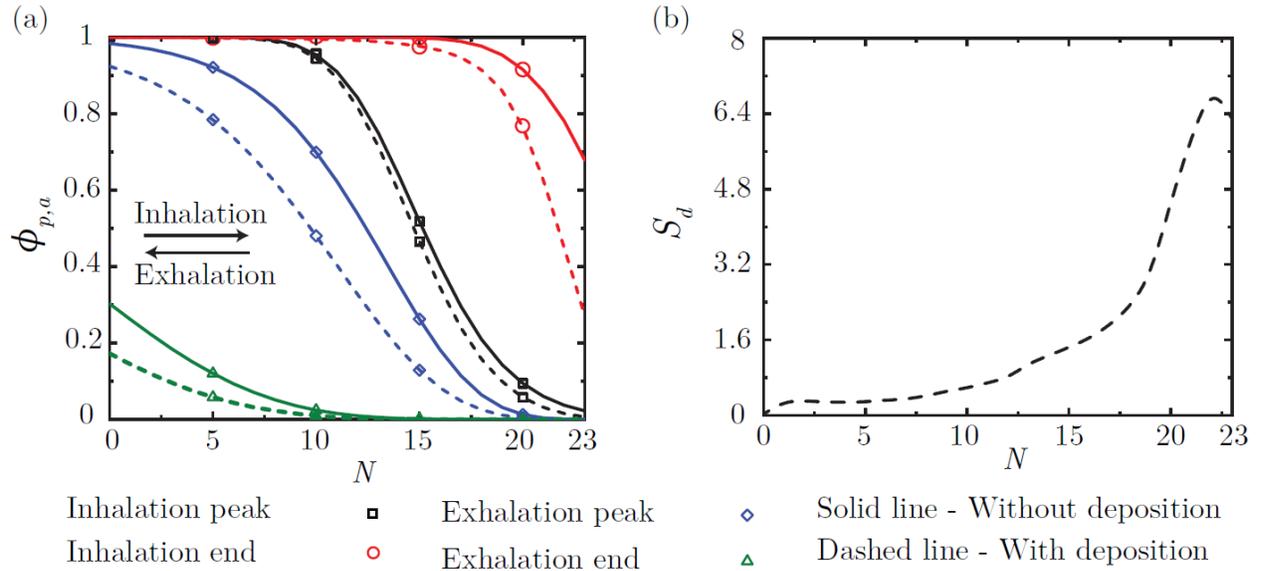

Figure 6: (a) Particle concentration ($\varphi_{p,a}$) within the lung at different instances of a single breathing cycle with and without deposition (b) Total droplet deposition ($S_d = \int\int L^0_{d} \varphi_{p,a} dV \, d\tau$) within the lung at the end of exposure. The results are shown for $Pe_{p,a} = 2.85 \times 10^{10}$, $St_a = 0.0095$ and $\tau_{exp} = 5$.

Particles are observed to deposit in the deep lung only when $Pe_{p,a}$ remains in the range of $2.37 \times 10^6 - 3.07 \times 10^{11}$ (see Fig. A6a). The reason being, at small $Pe_{p,a}$, the advection is not strong enough to carry the particles into the deep lung, whereas at large $Pe_{p,a}$ the particles deposit primarily in the upper airways due to the impaction mechanism [6, 14, 56, 69]. This range translates to particle diameters of 0.006 $\mu$m to 20 $\mu$m for normal breathing in a healthy individual (tidal volume of 1000 ml and $T_b = 4$ s). Within this range, deposition in the deep lung increase up to $Pe_{p,a} \sim 10^8$ and then decreases to reach a minimum when $4.29 \times 10^9 < Pe_{p,a} < 1.6 \times 10^{10}$ (particle diameters $\sim 0.4 - 1.2$ $\mu$m). Beyond this, another increase in deposition in the deep lung take place with a peak at $Pe_{p,a} = 9.03 \times 10^{10}$ (particle diameter of 3 $\mu$m) [6].

A longer breathing time period (lower $St_a$) results in *deeper* breaths and lead to greater volume of air (and hence, particles) being inhaled, keeping all other parameters same and vice-versa. This increases the probability of the inhaled particle reaching the deep lung. Consequently, the deposition pattern also shifts towards the deep lung with decrease in $St_a$ (see Fig. A6c-d). The fraction of particles deposited in the deep lung are, as such, observed to increase as $St_a$ decreases (see Fig. A6b) [6].

## 3. Mucus fluid dynamics

### 3.1. Mucus clearance

The mass and momentum balance equations of the mucus layer within the lung can be written, based on existing literature [15, 16, 70, 71], as

*Mass balance:*

$$\frac{\partial A_{m,N}}{\partial t} + \frac{\partial Q_m}{\partial x} = S_{m,N}.$$

(39)

*Momentum balance:*

$$\rho_m \frac{\partial}{\partial t}\left(\frac{Q_m}{A_{m,N}}\right) = -\frac{\partial p_a}{\partial x} + f'_c - f'_{visc},$$

(40)

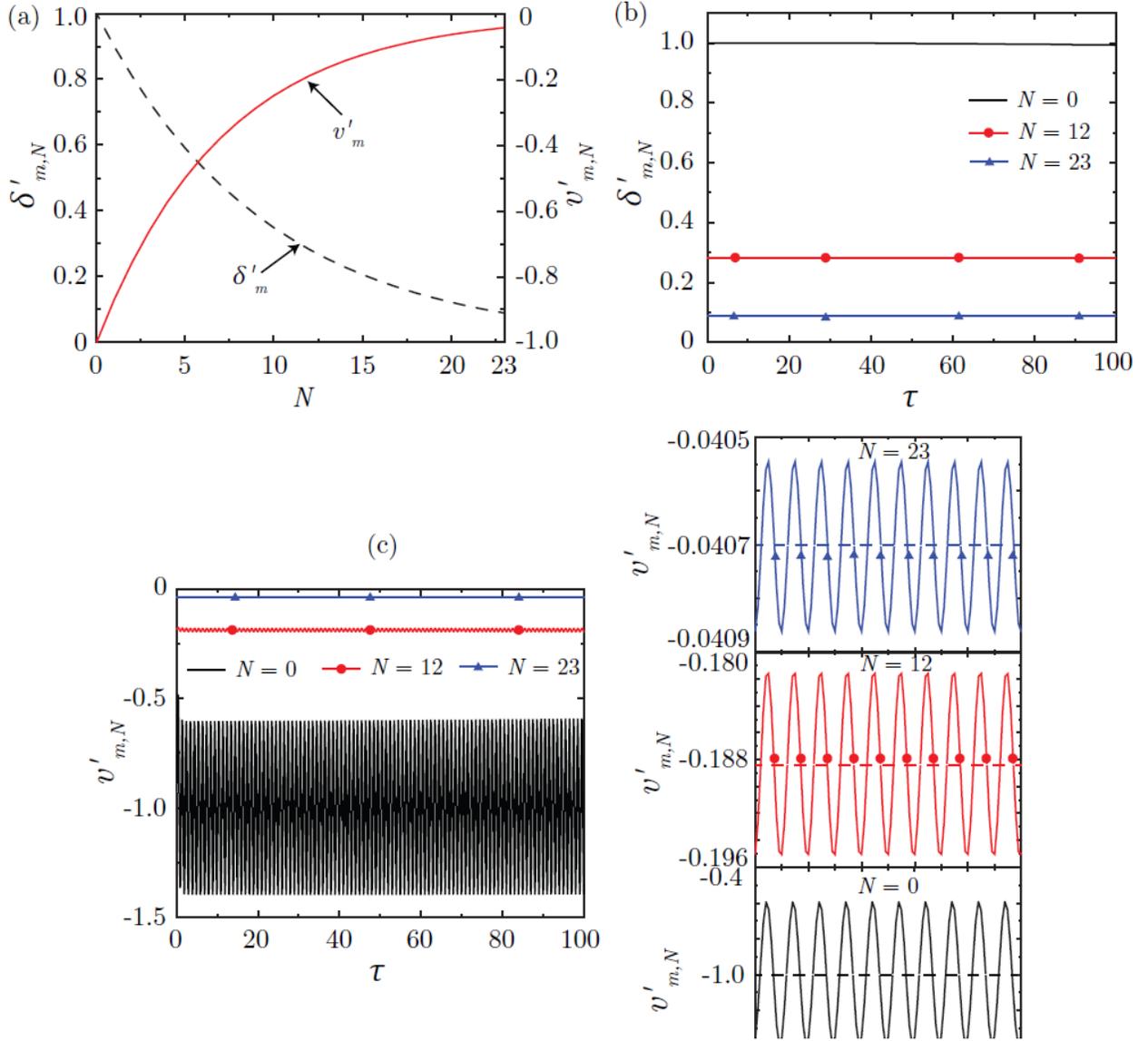

Figure 7: (a) Variation of steady-state mucus layer thickness ($\delta_{m,N}$) and mucus velocity ($v'_m$) along the lung generations (b) Temporal change in $\delta_{m,N}$ at three different lung generations ($N$=0,12,23) and (c) Temporal change in $v_{m,N}$ at three different 23). The results are shown considering $Re_m = 0.001, St_m = 359.7122, \alpha = 0.73, \beta = 0.71, P_0 = \sin(\omega\tau)$ $(2\beta^2)^N$.

where $f'_c$ and $f'_{visc}$ are the mucociliary propulsive force and the mucosal viscous resistive force per unit mucosal volume, respectively. The convective mucus transport term is neglected due to low Reynolds number flow in the thin viscous mucus layer. Note that the airway pressure gradient $\frac{\partial p_a}{\partial x}$ is assumed to be imposed in the mucus layer. This is similar to how pressure outside the boundary layer gets imposed within it. $A_{m,N}$ and $S_{m,N}$ are the area of the mucus layer and mass source of the mucus at the $N^{th}$ lung generation. $Q_m$ is the mucus volume flow rate and $\rho_m$ is the density of the mucus. These quantities are defined as

$$A_{m,N} = 2\pi R_N \delta_{m,N} 2^N; Q_m = A_{m,N} v_{m,N}; S_{m,N} = 2\pi R_N G_{m,N} 2^N; G_{m,N} = G_{m,0} \psi_g^N,$$

where $\delta_{m,N}$ is the thickness of the mucus layer at the $N^{th}$ lung generation and $v_m$ is the mucus velocity. $G_m$ represents the volumetric mucus source per unit area per unit time. Using these definitions and $H_N = \partial N / \partial x$, the mass balance equation (Eq. 39) is re-written in terms of $N$ as

$$\frac{\partial}{\partial t}\left(2\pi R_N \delta_{m,N} 2^N\right) + H_N \frac{\partial}{\partial N}\left(2\pi R_N \delta_{m,N} 2^N v_m\right) = 2\pi R_N G_{m,N} 2^N, \tag{41}$$

which can be simplified into

$$\frac{\partial}{\partial t}\left(\delta_{m,N}\right) - \frac{(1-\alpha)}{L_0 \alpha \ln(\alpha) R_N (2\alpha)^N} \frac{\partial}{\partial N}\left(R_N \delta_{m,N} 2^N v_m\right) = G_{m,N} \tag{42}$$

Similarly, the momentum balance equation (Eq. 40) can be converted in terms of lung generation number ($N$) using the above parameters and the converted equation is expressed as

$$\rho_m \frac{\partial}{\partial t}\left(\frac{Q_m}{A_{m,N}}\right) = \frac{8\pi\mu_a Q_a}{2^N A_0^2 \beta^{2N}} + \frac{k_c \mu_m h_c v_c N_{c0} N_c'}{\delta_{m,N}} - \frac{k_v \mu_m v_m}{\delta_{m,N}^2} \tag{43}$$

where

$$\frac{\partial p_a}{\partial x} = -\frac{8\pi\mu_a Q_a}{2^N A_0^2 \beta^{2N}},$$

$$f_c' = k_c(\mu_m h_c v_c N_{c0} N_c')\left(\frac{1}{\delta_{m,N}}\right),$$

$$f_{visc}' = k_v \left(\frac{\mu_m v_m}{\delta_{m,N}}\right)\left(\frac{1}{\delta_{m,N}}\right).$$

In the expressions above, $\mu_m$ is the mucosal viscosity, $h_c$ is the height of each cilia, $v_c$ is the cilia driving velocity scale, $N_{c0}$ is the number of cilia per unit surface area in generation 0, $N_c$ is the number of cilia in generation $N$, and $N_c' = N_c/N_{c0}$.

In the expression for $f_c'$, the term in the first parenthesis is the scale of the total mucociliary driving force, the term in the second parenthesis arises due to conversion to force per unit volume, and $k_c$ is the proportionality constant. Similarly, in the expression for $f_{visc}'$, the term in the first parenthesis is the scale of the viscous shear resistance to mucosal flow, the term in the second parenthesis arises due to conversion to force per unit volume, and $k_v$ is the proportionality constant.

In order to obtain the dimensionless form of Eqs. 42 and 43, we define the following

$$\tau = \frac{t}{T_b}, \delta_{m,N}' = \frac{\delta_{m,N}}{\Delta_{m,s}}, v_{m,N}' = \frac{v_{m,N}}{v_{m,s}}, \tag{44}$$

where $T_b$ is the breathing time scale. $\Delta_{m,s}$ and $v_{m,s}$ are the mucus layer thickness scale and the velocity scale, respectively. These scales are estimated based on steady-state scalings implied by mass and momentum balance equations. The dominant steady state scaling is between the last two terms in each of the Eqs. 42 and 43, which gives

$$\Delta_{m,s} = \sqrt{\frac{k_v G_{m,0} L_0}{k_c h_c v_c N_{c,0}}}, v_{m,s} = \frac{G_{m,0} L_0}{\Delta_{m,s}} \tag{45}$$

The dimensionless equations, thus, obtained can be written as

$$St_m \frac{\partial}{\partial t}\left(\delta_{m,N}'\right) - \left[\frac{(1-\alpha)}{\alpha \ln(\alpha)(2\alpha^2)^N} \frac{\partial}{\partial N}\left(\delta_{m,N}' v_{m,N}'(2\alpha)^N\right)\right] = \psi_g^N \tag{46}$$

$$Re_m St_m \frac{\partial v_{m,N}'}{\partial t} = P_Q - \left(\frac{N_c'}{\delta_{m,N}'}\right) - \left(\frac{1}{\delta_{m,N}'}\right)\frac{v_{m,N}'}{\delta_{m,N}'}, \tag{47}$$

where $St_m\left(=\frac{T_m}{T_b}=\frac{L_0}{v_{m,s}T_b}\right)$ and $Re_m\left(=\frac{\rho_m G_{m,0}\Delta_{m,s}}{k_v \mu_m}\right)$ are the Strouhal number and Reynolds number for the mucus layer, respectively. The parameter $P_Q$ is defined as

$$\left(\frac{8\pi\mu_a Q_a \Delta_{m,s}^2}{(2\beta^2)^N A_o^2 k_v \mu_m v_{m,s}}\right)$$

Equations 46 and 47 are solved in a coupled manner to obtain the spatio-temporal change in $\delta'_m$ and $v'_m$ within the lung. It is initially assumed that $\delta'_m$ and $v'_m$ follow a power-law spatial change, similar to the assumptions made for airway length and area (Eqs. 1), in accordance with the following equations

$$\delta'_{m,N} = \delta'_{m,0}\zeta^N,$$
(48)

$$v'_{m,N} = v'_{m,0}\varepsilon^N,$$
(49)

where $\delta_m{}^0{}_0$ and $v_m{}^0{}_0$ are the dimensionless mucus thickness and mucus velocity, respectively, at $N = 0$. The magnitudes of $\zeta$ and $\varepsilon$ are considered to be 0.9 and 0.87, respectively, based on reported data [16]. The spatial change due to the above assumption is represented in Fig. 7a. Note that the mucus velocity has a negative magnitude since mucus flow is in a direction opposite to increasing depth of the lung. $N_c{}^0$ is assumed to vary as $\psi_n{}^N$ where $\psi_n = \dfrac{\varepsilon}{\zeta}$. $\psi_g$ is defined as $\dfrac{\varepsilon\zeta}{\alpha}$.

Solution of the equations indicate that the dimensionless mucus layer thickness ($\delta'_m$) remain almost invariant with time (see Fig. 7b). However, the dimensionless mucus velocity ($v'_m$) undergoes a timeperiodic change about a mean value (see Fig. 7b), which is in phase with the breathing time-period, after an initial period of transience. Airflow during inhalation resists the mucus transport and hence, mucus velocity reduces during inhalation. The opposite happens during exhalation due to co-current flow of air and mucus. Similar characteristics are observed at all lung generations although the mucus velocity becomes negligible in the deep lung. The transience period is, however, very short and the time-periodic state is reached in a few breathing cycles. Such transience time-period is very small when compared to the time-scale of particle clearance through mucus transport. Hence, it can be safely neglected when modelling particle clearance. This, coupled with the time-periodic nature of mucus velocity, allows us to assume a steady-state magnitude of mucus velocity within the lung when modelling particle clearance (see Section 3.2).

### 3.2. Clearance of deposited particles

The one-dimensional transport equation for the particles deposited in the lung mucus, by the various deposition mechanisms (see Section 2.3), is formulated considering mucociliary transport and diffusion of the deposited particles in the mucus. It can be expressed in a similar manner to Eq. 25 as follows [6, 36]-

$$\frac{\partial(A_m c_{p,m})}{\partial t} + H\frac{\partial(Q_m c_{p,m})}{\partial N} = H\frac{\partial}{\partial N}\left(A_m D_{p,m} H \frac{\partial c_{p,m}}{\partial N}\right) + L_D c_{p,a}\phi_l,$$
(50)

where $c_{p,m}$, $A_m$, $Q_m$, and $D_{p,m}$ are deposited particle concentration in the mucus, cross-sectional area of the mucus layer at a particular lung generation, volume flow rate of mucociliary transport, and particle diffusivity in the mucus, respectively. It is assumed that the mucus velocity ($v_m$) and hence, $Q_m$ remains time-invariant (see Section 3.1). $\phi_l$ is defined as the quantity of particles, which may get deposited, per unit quantity of particles inhaled. The term $L_D c_{p,a}\phi_l$ takes into account the particles introduced into the mucus due to deposition. Eq. 50 is converted to a dimensionless form (Eq. 52) using scaling defined in Eq. 51 below [6, 36]

$$\tau = \frac{t}{T_b}, \phi_{p,m} = \frac{c_{p,m}}{c_{p,m,0}}, c_{p,m,0} = \phi_l c_{p,a,0}\frac{A_0}{A_{m,0}}, T_m = \frac{L_0}{|v_{m,0}|}, \quad St_m = \frac{T_m}{T_b}, Pe_{p,m} = \frac{|v_{m,0}|L_0}{D_{p,m}}, D_p = \frac{k_B T}{3\pi\mu_m d_p}.$$
(51)

$$|Pe_{p,m}|(2\alpha\zeta\sqrt{\beta})^N St_m \frac{\partial\phi_{p,m}}{\partial\tau} = \frac{\partial F_{p,m}}{\partial N} + \left(L'_D \frac{D_{p,a}}{D_{p,m}}\phi_{p,a}\right),$$
(52)

where $\varphi_{p,m}$, $Pe_{p,m}$, and $St_m$ are the dimensionless particle concentration in the mucus, particle Peclet number, and mucus layer Strouhal number, respectively. Also note that $Pe_{p,m}$ refers to the particle Peclet number at $N = 0$ only. $T_m$ denotes the time-scale for mucociliary transport. $D_{p,m}$ is estimated using the Stokes-Einstein relation, where $\mu_m$ and $d_p$ are the viscosity of the mucus and the particle diameter, respectively [72]. The last term on the right hand side of Eq. 52 is the dimensionless particle source due to deposition. The flux term in Eq. 52 is represented as

$$F_{p,m} = \left[ \left( \left( \frac{2\zeta\sqrt{\beta}}{\alpha} \right)^N \left( \frac{1-\alpha}{\alpha \ln(\alpha)} \right)^2 \frac{\partial \phi_{p,m}}{\partial N} \right) - \left( |Pe_{p,m}|(2\varepsilon\zeta\sqrt{\beta})^N \phi_{p,m} \right) \right]$$

(53)

The particles that are inhaled into the lung are deposited in the respiratory mucus during breathing by various mechanisms (see Section 2.3) [13, 18, 19]. The deposited particles diffuse in the mucus layer and are also simultaneously transported upstream via mucociliary advection [73]. However, mucociliary advection is appreciable only in the conducting airways of the lung ($N < 18$) and is negligible in the deep lung ($N \geq 18$) [16]. Results show that the particle concentration in the mucus ($\varphi_{p,m}$), at the end of the exposure duration ($\tau = 5$), qualitatively follows particle deposition $S_d$ (see Figs. A8a and 6b).

Particles deposited in the conducting airways are transported upstream towards the mouth ($N = 0$) due to mucociliary advection. This results in higher $\varphi_{p,m}$ in the upper airways (lower $N$), as time progresses, primarily due to smaller mucus volume in the upper airways. Eventually, the particles are washed out of the respiratory tract (see Fig A8a) along with the mucus. The temporal change in $\varphi_{p,m}$ at the mouth (inset Fig. A8a) also corroborates this conclusion. In contrast, the particles deposited in the deeper generations ($N \geq 18$) is not subjected to mucociliary transport. Therefore, $\varphi_{p,m}$ in the deep lung undergoes a gradual change due to weak diffusive transport (see inset Fig. A8a). As such, particles deposited in the deep lung can persist for a much longer time as compared to those deposited in the upper airways [6, 36].

The washout of the deposited particles from the lung is observed to be dependent on $St_m$. $St_m$ is defined as the ratio of the mucociliary advection time scale to the breathing time scale (see Eq. 51). A longer $T_b$ relative to the mucociliary advection time scale leads to a smaller $St_m$. This implies greater advective clearance of the mucus in a breathing cycle. Thus, a larger amount of deposited particles are washed out along with the mucus in case of smaller $St_m$ (see Fig. A8b). However, the enhanced particle clearance is observed to remain limited to the upper airways only and does not influence particle clearance from the deep lung [6, 36].

## 4. Pathogen infection dynamics

Pathogens are introduced into the LRT through the carrier particles (or droplets) which are transported in the respiratory tract along with airflow. The carrier particles (or droplets) may be inhaled or formed in-vivo through aerosolization of respiratory mucosa [8, 36]. The carrier particles (or droplets) deposit in the mucosa of the LRT (see Section 2.3), thereby, seeding an infection. The deposited particles (or droplets) are also simultaneously cleared through advective-diffusive transport (see Section 3.2). The infection once seeded - may grow, spread and decay depending on various pathogen-specific parameters and the host's immune responses. Modeling the dynamics of a pathogen infection within the respiratory tract, thus, requires simultaneous consideration of all the above mechanisms.

### 4.1. Pathogen transport in mucus

The one-dimensional pathogen transport equation within respiratory mucosa, thus formulated, is represented in its dimensionless form as [36]

$$Pe_{pt,m}(2\alpha\zeta\sqrt{\beta})^N St_m \frac{\partial \phi_{pt,m}}{\partial \tau} = \frac{\partial}{\partial N} \left[ \left( \left( \frac{2\zeta\sqrt{\beta}}{\alpha} \right)^N \left( \frac{1-\alpha}{\alpha \ln(\alpha)} \right)^2 \frac{\partial \phi_{pt,m}}{\partial N} \right) - \left( Pe_{pt,m}(2\varepsilon\zeta\sqrt{\beta})^N \phi_{pt,m} \right) \right. $$
$$\left. + \left( L'_D \frac{D_{p,a}}{D_{pt,m}} \phi_{p,a} \right) + (2\alpha\zeta\sqrt{\beta})^N \left( p_0 I - c_l \phi_{pt,m} \right), \right.$$

(54

) where $\varphi_{pt,m}$, $Pe_{pt,m}$ and $St_m$ are the dimensionless pathogen concentration in mucus, pathogen Peclet number and mucus Strouhal number, respectively. Equation 54 is similar to Eq. 52 except the last term on the right hand side of Eq. 54 which accounts for the pathogen-specific infection kinetics using a modified target-cell limited model [36, 37, 74].

The modified target-cell limited model (Eqs. 55-57) assumes that the target cells are infected depending on the pathogen-specific infection rate and the local pathogen concentration (see Eq. 55). The infected target cells remain in an eclipse phase for a certain duration before becoming infectious (Eq. 56) after which they produce new pathogens for a certain time-span before undergoing apoptosis (Eq. 57). The corresponding equations are expressed as follows -

$$\frac{\partial T}{\partial \tau} = -I_r T \phi_{pt,m},$$

(55)

$$\frac{\partial E}{\partial \tau} = (I_r T \phi_{pt,m}) - \left(\frac{1}{\tau_E}E\right),$$

(56)

$$\frac{\partial I}{\partial \tau} = \left(\frac{1}{\tau_E}E\right) - \left(\frac{1}{\tau_I}I\right).$$

(57)

$T$, $E$ and $I$ in the above equations represent the fraction of uninfected target cells, infected cells in the eclipse phase and infectious cells, respectively. $I_r$ denotes the dimensionless infection rate. $\tau_E$ and $\tau_I$ are the dimensionless time-periods of the eclipse phase and infectious phase, respectively. $p_0$ and $c_l$ in Eq. 54 denote the dimensionless pathogen production rate of new pathogens from the infectious cells and the dimensionless clearance rate of pathogens due to various non-specific clearance mechanisms, respectively [37, 74]. The relevant parameters are defined as [36]

$$p_0 = \frac{L_0^2}{D_{pt,m}}\frac{p}{c_{pt,m,0}}, c_l = \frac{L_0^2}{D_{pt,m}}c, I_r = \beta c_{pt,m,0} T_b, \tau_E = \frac{T_E}{T_b}, \tau_I = \frac{T_I}{T_b},$$

(58)

where $c_{pt,m}$ is the dimensional pathogen concentration and $c_{pt,m,0}$ is the initial $c_{pt,m}$ at $N = 0$. $p$, $c_l$ and $\beta$ are the dimensional rate of pathogen replication, virus clearance and infection, respectively. $T_E$ and $T_I$ are the time-scales for the eclipse phase and the infectious phase, respectively.

### 4.2. Immune Response to an infection

A simplified immune response model [6, 37], coupled with the infection kinetics model, is utilised to understand the impact of human body's immune response to pathogen infections in the respiratory tract. The model considers the effects of interferons, antibodies and cytotoxic $T$-lymphocytes. The specific mathematical models for the individual immune responses are discussed in the following sections [36].

### 4.2.1. Interferon response

Interferons are assumed to affect infection progression by attenuating replication of the pathogens. The pathogen replication rate ($p_0'$) is, thus, reduced in presence of interferons and is determined as

$$p_0' = \left(1 - \frac{F}{F+f}\right)p_0,$$

(59)

where $p_0$ is the pathogen replication rate in absence of interferons (see Eq. 54) and $f$ is the interferon fraction required to halve the replication rate. $F$ is the interferon fraction present in the body (relative to the maximum amount interferons that may be present) and it is assumed to vary with time as

$$F = \frac{2}{e^{-\lambda_{g,i}(\tau-\tau_{p,i})} + e^{\lambda_{d,i}(\tau-\tau_{p,i})}},$$

(60)

where $\lambda_{g,i}$ and $\lambda_{d,i}$ are the dimensionless rates of interferon growth and decay, respectively. $\tau_{p,i}$ is the dimensionless time at which interferon fraction reaches its maxima.

### 4.2.2. Antibody response

Antibodies act by neutralising the pathogens present in the body. The pathogen clearance rate ($c_l$ in Eq. 54) is, thus, enhanced due to antibodies and is determined as

$$c_l' = c_l + k_A' Ab, \tag{61}$$

where $c^0_l$ is the enhanced clearance rate due to antibodies and $k_A^0$ is the dimensionless binding affinity of antibodies to the pathogens. $Ab$ is the antibody fraction present in the body and it varies with time as

$$Ab = \frac{1}{1 + \left(\frac{1}{Ab_0} - 1\right)e^{-\lambda_{g,a}\tau}}, \tag{62}$$

where $\lambda_{g,a}$ is the dimensionless antibody growth rate and $Ab_0$ is the initial antibody fraction present in the body.

### 4.2.3. T−lymphocyte response

The cytotoxic $T$−lymphocytes act by directly attacking the infected cells. Equations 56 and 57 are, thus, modified when $T$−lymphocytes are present as

$$\frac{\partial E}{\partial \tau} = (I_r T \phi_{pt,m}) - \left(\frac{1}{\tau_E}E\right) - (k_c' T_l)E, \tag{63}$$

$$\frac{\partial I}{\partial \tau} = \left(\frac{1}{\tau_E}E\right) - \left(\frac{1}{\tau_I}I\right) - (k_c' T_l)I, \tag{64}$$

where $k_c^0$ represents the dimensionless rate of infected cells neutralisation by the $T$−lymphocytes. $T_l$ is the fractional amount of $T$−lymphocytes present at any time and is determined as

$$T_l = \frac{2}{e^{-\lambda_{g,t}(\tau-\tau_{p,t})} + e^{\lambda_{d,t}(\tau-\tau_{p,t})}}, \tag{65}$$

where $\lambda_{g,t}$ and $\lambda_{d,t}$ are the dimensionless rates of $T$−lymphocyte growth and decay, respectively. $\tau_{p,t}$ is the dimensionless time at which amount of $T$−lymphocytes present in the body reaches its maxima.

### 4.3. Infection progression

The pathogen transport model (Sections 4.1 and 4.2) coupled with the particle transport and deposition model (Section 2.3) have been utilised to study the dynamics and progression of a SARS-CoV-2 infection in a human LRT [36]. It is assumed during this analysis that virus-loaded droplets are present in the naso-pharyngeal region of the respiratory tract. These droplets may be a combination of droplets inhaled from the environment (which have not deposited in the upper respiratory tract) and those formed due to aerosolization of nasopharyngeal mucosa, and are inhaled into the trachea during breathing. No other source of droplets are considered.

It is observed from the analysis that pathogen (SARS-CoV-2) concentration in mucus ($\varphi_{pt,m}$), at the end of the inhalation, is qualitatively similar to droplet deposition characteristics. This is due to a significantly shorter time-scale of droplet deposition, as compared to pathogen infection and clearance from the LRT [36]. As droplet inhalation stops, mucociliary clearance transports the pathogens (viruses) deposited in the upper airways of the LRT ($N < 18$) upstream towards the trachea ($N = 0$). This can be inferred from the spatial change in ($\varphi_{pt,m}$) with time (see Fig. A9a). This mechanism washes out the pathogens (viruses) from the upper airways as long as washout dominates over pathogen (virus) replication. Once pathogen (virus) replication starts to dominate, $\varphi_{pt,m}$ is observed to again increase with time indicating infection growth within the LRT (Fig. A9a-b).

In contrast to pathogens (viruses) deposited in airways, pathogens (viruses) deposited in the deep lung ($N \geq 18$) are transported only through diffusion since mucociliary clearance is negligible in the deep lung. The dynamics of $\varphi_{pt,m}$ in the deep lung is, thus, governed by the weaker diffusive transport (compared to mucociliary clearance) and pathogen (virus) kinetics only. As a result, deep lung deposition of pathogens (viruses) leads to longer persistence times. This increases the probability of an infection becoming severe causing serious diseases (pneumonia, acute respiratory distress syndrome etc.). In addition, the thin surfactant layer lining the deep lung increases the possibility of the pathogens entering the blood stream.

The LRT infection is observed to grow as long as pathogen (virus) replication dominates over pathogen (virus) clearance. Once clearance becomes stronger, infection starts to recede ($\varphi_{pt,m}$ decreases) with time (see

Fig. A9b). This results in a peak infectious state beyond which infection starts to resolve. The peak infectious state and the infection resolution time is observed to be dependent on several fluid dynamics, physiological as well as infection parameters (see Chakravarty et al. [36] for details). The analysis reveals that the major impact is due to immune responses, paticularly antibodies (Fig. A10a) and cytotoxic Tlymphocytes (Fig. A10b), which reinforces the critical role played by vaccination in preventing infection severity.

## 5. Discussion

### 5.1. Utility and limitations of 1D models

High resolution computational simulations of the complete respiratory tract becomes difficult due to the geometrical complexity of the respiratory tract [16]. Analyses have shown that the transport phenomena within the respiratory tract has a time-periodic nature which enables 1D approximations with reasonable accuracy [44]. Simplified 1D computational models would be reasonable when the focus is to capture the key trends of droplet/particle/pathogen transport and deposition for the complete respiratory tract. While such simplified models cannot account for the effects of heterogeneity in the lung, it is a tractable approach and can help understand some of the key trends dependent on the entire respiratory tract (see Sections 2 and 3). Investigation of pulmonary drug delivery using a simplified 1D model (see Sections 2.3 and 3.2) has provided information regarding spatio-temporal drug deposition characteristics throughout the respiratory tract and also suggested ways of enhancing drug delivery efficacy [6]. The pathogen infection model (see Section 4.1) has been utilised to study the spatio-temporal evolution of SARS-CoV-2 infection in a human respiratory tract and provided useful information on the role played by various physiological and fluid dynamic parameters on infection characteristics [36]. Similar investigations can be carried out considering other viral infections as well.

The 1D single trumpet model does not fully resolve complicated flow patterns in the URT or the effect of heterogeneity. However, a network of dichotomously branched heterogeneous lung could be modeled by approximating each branch as a 1D tube with similar governing equations as the trumpet model. This is reported before [63, 75] and also summarized in "asymmetric lung modeling" in 5.3.3, below.

### 5.2. Dimensionless numbers determining lung function

Non-dimensional numbers give an understanding of the dominant physics in the governing equations and therefore the lung function. This is evident from the equations and results summarized in this article in the previous sections. Table 2 summarizes the dimensionless numbers that govern the various lung functions - airflow, gas exchange, particle deposition and clearance, and infection dynamics - and their typical physiological magnitudes. The magnitude of infection dynamics parameters are relevant for a typical SARSCoV-2 infection [6] and are subject to variations while modeling other respiratory infections. Finally, the non-dimensional numbers listed in Table 2 could form the basis of constructing a "virtual disease landscape" discussed in Outlook, below.

### 5.3. Optimal branching of the lung

One of the questions of interest is whether the lung morphology has developed to be optimal for certain function(s). And if so how would deviations from standard morphology potentially lead to pathologies? The primary function of the lung is gas exchange. Thus, its morphology might be optimal for air flow and/or gas exchange. The former has been reported in literature (see Section 5.3.1 below) and some insights on the latter are discussed below (see Section 5.3.2). Asymmetry of the branching structure of the lung may also affect optimality (see Section 5.3.3).

While particle deposition and clearance in the lung has been extensively studied, more research is needed to determine whether the lung morphology is optimal to minimize particle deposition from inhaled air and to maximize particle removal via mucus clearance. We note that particle deposition is different from gas exchange although the governing equations are similar (see above). This is because, particles are not absorbed easily into the bloodstream like, say, oxygen because they are much bigger in size. Instead, particles deposit in the mucosal layer and mucus clearance transports the particles out of the lung.

Table 2: Non-dimensional numbers and their physiological magnitudes encountered during airflow, gas exchange, particle deposition and clearance, and infections in the respiratory tract. These numbers are defined in the respective sections of this article. *The magnitudes corresponding to infection dynamics are relevant for a typical SARS-CoV-2 infection.

| | Dimensionless Parameter | Physiological magnitude |
|---|---|---|
| **Airflow** | $Re$ | $\sim 10^3 - 6 \times 10^{-3}$ |
| | $Wo$ | $\sim 6 - 0.1$ |
| | $\alpha$ | $0.73$ |
| | $\beta$ | $0.71$ |
| | $r$ | $0 - 0.5$ |
| **Gas exchange** | $Pe_g$ | $\sim 1000 - 65000$ |
| | $St_a$ | $\sim 0.001 - 0.1 \sim$ |
| | $Z$ | $500 - 2500$ |
| | | $Z_{cr} = 2295.827$ |
| | $\tau_{exp}$ | $5$ |
| **Particle deposition &clearance** | $Pe_{p,a}$ | $\sim 10^7 - 10^{12}$ |
| | $Pe_{p,m}$ | $\sim 10^7 - 10^8$ |
| | $St_a$ | $\sim 0.001 - 0.1$ |
| | $St_m$ | $\sim 100 - 1500$ |
| | $\tau_{exp}$ | $5$ |
| | $\zeta$ | $0.9$ |
| | $\epsilon$ | $0.87$ |
| **Infection dynamics*** | $p_0$ | $3.8 \times 10^{16} - 5.62 \times 10^{17}$ |
| | $c_i f$ | $0 - 3.79 \times 10^8$ |
| | $\tau_{p,i}$ | $0.2 - 1$ |
| | $\lambda_{g,i}\ \lambda_{d,i}$ | $2.16 \times 10^4 - 19.44 \times 10^4$ |
| | $k_{A0}$ | $9.26 \times 10^{-5}$ |
| | $Ab_0$ | $4.63 \times 10^{-5}$ |
| | $\lambda_{g,a}\ k_{C0}$ | $0 - 9.1 \times 10^8$ |
| | | $0.001 - 0.005$ |
| | | $3.472 \times 10^{-5}$ |
| | | $0 - 10^{-3}$ |
| | $\tau_{p,t}$ | $8.64 \times 10^4 - 34.56 \times 10^4$ |
| | $\lambda_{g,t}$ | $9.26 \times 10^{-5}$ |
| | $\lambda_{d,t}$ | $4.63 \times 10^{-6}$ |

### 5.3.1. Optimization based on air flow

There are at least two ways to investigate whether the lung morphology is optimal to minimize resistance to air flow. The first approach involves an analysis where the total length and volume of different dichotomous trees being compared are held fixed. Then $\alpha$ and $\beta$ can be optimized such that the total flow resistance is minimized. It has been that the physiological values of $\alpha$ and $\beta$ are close to the optimal values [70, 76].

A second alternate approach is to fix the dimensions of the zeroth generation (trachea) and then keep adding dichotomous generations. Clearly, the length and volume will keep increasing as one considers scenarios with increasing number of generations. Hence, to compare across these scenarios with different number of generations, one can do a comparison of a dimensionless resistance normalized by the resistance of straight tube of the same length and volume as a particular tree network being analyzed. This is similar to the concept of drag coefficient for flow over objects. To that end, consider a straight tube of the same length $x$ and the same total volume as a trumpet model with $N$ generations. Assuming parabolic flow in this equivalent straight tube, its flow resistance $R_s$ is given by

$$R_s = R_0 \left( \frac{1 - \alpha^{N+1}}{1 - \alpha} \right)^3 \left( \frac{1 - 2\alpha\beta}{1 - (2\alpha\beta)^{N+1}} \right)^2.$$

(66)

The dimensionless ratio, $R^+$, of $R_s$ to the resistance of a trumpet model with $N$ generations is given by

$$R^+ = \frac{R_s}{R_{tN}} = R_0 \left( \frac{1 - \alpha^{N+1}}{1 - \alpha} \right)^3 \left( \frac{1 - 2\alpha\beta}{1 - (2\alpha\beta)^{N+1}} \right)^2 \left( \frac{1 - \left( \frac{\alpha}{2\beta^2} \right)}{1 - \left( \frac{\alpha}{2\beta^2} \right)^{N+1}} \right).$$

(67)

Maximizing $R^+$ (i.e. minimizing the relative value of $R_{tN}$) leads to the optimal values of $\alpha$ and $\beta$. It follows from Eq. 68 that the optimal value of $\alpha = 0.78$ for $N = 23$ and $\beta = \alpha^2$ [77], which is close to the physiological value.

*5.3.2. Optimization based on gas exchange*

Optimization of lung structure for gas exchange is another consideration [78]. Similar to the second approach for flow resistance-based optimization one can optimize the resistance to the impalement of gas into the deep lung. To that end, consider the total resistance $R_{ct}$ to the flux of concentration. We define $R_{ct}$ as the concentration difference per unit concentration flux. We compare the trumpet configuration concentration resistance $R_{ct}$ to the concentration resistance $R_{cs}$ in a straight tube of the same length and volume. Following theoretical solutions as presented earlier we get

$$R_c^+ = \frac{R_{cs}}{R_{ct}} = \left( \frac{1 - \alpha^{N+1}}{1 - \alpha} \right)^3 \left( \frac{1 - 2\alpha^3}{1 - (2\alpha^3)^{N+1}} \right)^3 (2\alpha^3)^N \left( \frac{1 - \exp(-Pe_s)}{1 - \exp(Pe_{g,pl}(\delta^N - 1))} \right),$$

(68)

where $\beta = \alpha^2$ has been used and $R_c^+$ is the ratio of concentration resistances discussed above. $Pe_{st}$ is the Peclet number of the straight channel of same length, volume, and pressure difference as the trumpet channel. The concentration at $N = 23$ is taken to be zero for all practical purposes and the inlet concentration is a specified constant value. Eq. 68 shows that $R_c^+$ is maximized for $\alpha > 2/3$, once again in the physiologically relevant range.

*5.3.3. Asymmetry and optimality*

Although the aforementioned optimality results assumed symmetric branching, those results may be regarded to be relevant in the "average" sense. An important feature of airway branching structure is also its asymmetry. While a majority of human airway models have assumed a simplified morphology of the lung by considering symmetric bifurcations [12], careful analysis of morphometric measurements have shown that the bifurcations are in fact asymmetrical in nature [52]. Surprisingly, there lies a consistency in the degree of asymmetry across all generations, although it varies from species to species [79]. Studies on investigating the optimal degree of asymmetry have led to interesting results [70, 75, 80–82]. Mauroy et al. [70] have shown that a mechanically optimal lung is vulnerable to broncho-constriction. Florens et al. [82] showed how oxygenation times start decreasing sharply as the degree of asymmetry is increased beyond a critical point.

Through deterministic asymmetric multi-path models of the bronchial trees, Kundu and Panchagnula [75] studied the optimality of the lung as a multi-functional organ. Their findings suggest that the number of terminal bronchioles, which is correlated to the gas exchange surface area, is maximized at symmetry (see Fig. A11a). The volume occupied by the conducting airways is minimized at symmetry, suggesting the most compact design (see Fig. A11b). The viscous resistance to air-flow, which should be as low as possible in an optimal lung, is also minimized for a symmetrically branched bronchial tree (see Fig. A11c). Breathing, the primary function of the lung, is thus optimized for a symmetrically branched bronchial tree based on these three parameters. Therefore, perfectly symmetric bifurcations appear to be the best design for the airways.

However, airway structure in the respiratory system is inherently asymmetric. This is necessitated due to the role played by airway asymmetry in particle filtration - the secondary function of the lung. As aerosolladen air is inhaled, the particles get deposited in the airways through three main physical mechanisms - diffusion, impaction and sedimentation [6, 36, 75]. Kundu and Panchagnula [75] have shown that for a finitely asymmetric bronchial tree, the particles get maximally deposited in the non-terminal branches of the lung, thereby protecting the deep lung from getting exposed to foreign particles (see Fig. A11d). The degree of asymmetry which theoretically maximizes this particle filtration efficiency is very close to that measured in human lung [79].

Thus, although symmetric branchings are the most optimal for maximizing gas exchange surface area, volume occupied and fluid dynamic resistance, asymmetry can maximize particle filtration by enhancing tracheo-bronchial deposition. This shift from mechanical optimality can enhance the protective mechanism of the airways.

### 5.4. Transition between convection and diffusion dominant regions

The concentration profiles during inhalation in the figures in the previous sections (e.g. Fig. A1) show that in the upper generations of the lung the gas transport (e.g. of oxygen) is convection dominated. The concentration profile appears almost constant; this is called the "conductive" region of the lung. In the deeper generations, the concentration profile is diffusion dominated. It is characterised by lower concentrations with very small gradients. The concentration profile shows a distinct transition between these two regions. The location of this transition can be theoretically estimated in the limiting case of steady state profile as discussed below.

It follows from Eq. 15 that

$$\left(Q - D_g \frac{\partial A_x}{\partial x}\right)\frac{\partial c_{g,x}}{\partial x} - A_x D_g \frac{\partial^2 c_{g,x}}{\partial x^2} = 0. \tag{69}$$

The last term is the diffusion-like term with an effective diffusion coefficient $A_x D_g$ that depends not only on gas diffusion $D_g$ but also on the cross-sectional area $A_x$ of the lung. Thus, increasing area of the lung enhances the effective diffusive transport in deep lung.

The first two terms in Eq. 69 are convection-like terms. Of these, the term involving $Q$ is the familiar flowbased convective transport term. The second term involving $\frac{\partial A_x}{\partial x}$ is present only because $A_x$ is not constant in the lung. This term behaves like a pseudo-convective term with an effect that is equivalent to a flow from deeper generations to the mouth (note: $\frac{\partial A_x}{\partial x} > 0$), i.e., like a reverse flow during inhalation. This has an effect of making the concentration profile look like a convection-dominated flow (near constant concentration profile) but in the reverse direction even in the so-called diffusion-dominated region (see Fig. A1). Thus, it should be expected that the convection-like term based on $Q$ would dominate the upper generations of the lung ("conductive region") while the convection-like term based on $A_x$ would dominate the deeper generations of the lung ("diffusive region"). The transition location is influenced by a competition between these two mechanisms, as shown below.

For the profile in the transition region, there will be an inflection point (see Fig. A1). Consequently, at the inflection point in the plot of $c_{g,x}$ vs. $x$, we have $\frac{\partial^2 c_{g,x}}{\partial x^2} = 0$ without $\frac{\partial c_{g,x}}{\partial x}$ becoming zero. This is possible when [83]

$$Q = D_g \frac{\partial A_x}{\partial x} = D_g \frac{\partial N}{\partial x} \frac{\partial A_N}{\partial N}. \tag{70}$$

Substituting relevant expression and solving for the generation number $N_I$ at the inflection point for a given constant flow rate $Q$, we get

$$N_I = \frac{\ln\left[-\frac{1}{Pe_g}\left(\frac{1-\alpha}{\alpha}\right)\frac{\ln(2\beta)}{\ln\alpha}\right]}{\ln\left(\frac{\alpha}{2\beta}\right)}. \tag{71}$$

Figure 8 shows the change in location of the transition zone (in terms of $N_I$) for a range of $Pe_g$. It can be observed that the transition zone progressively moves inwards into the lung with increase in convection (larger $Pe_g$). This can be corroborated with the gas concentration profiles shown in Fig. 3. Furthermore, the breadth of the transition region can also be estimated. We note that around the inflection point the convection and diffusion terms must be of the same order

$$Qc_{g,x} \sim A_x D_g \frac{\partial c_{g,x}}{\partial x}.$$

(72)

Inserting scales in the equation above, we get that the width $\delta_I$ of the transition region around the inflection point should be such that

$$Pe_I = \frac{Q\delta_I}{A_I D} \sim \mathcal{O}(1).$$

(73)

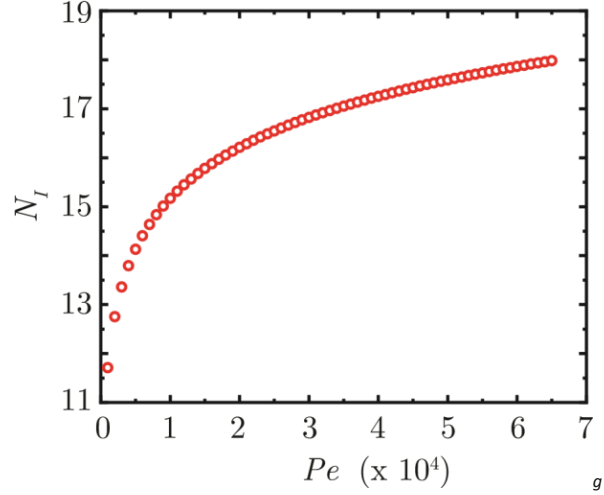

Figure 8: Change in position of the inflection point ($N_I$) with $Pe_g$. The entire physiological range of $Pe_g$ (based on gas diffusivity and flow rate) is considered.

This gives an estimate for $\delta_I$.

### 5.5. Nitrogen washout

Single or multiple-breadth washout (SBW or MBW) of nitrogen ($N_2$) is one of the tests used to quantify the degree of heterogeneity in the ventilation of the lung. Since $N_2$ is present in the air there is an equilibrium concentration in the deep lung during regular breathing. During the test, the subject breathes pure oxygen ($O_2$) so that $N_2$ starts getting washed out of the lung. During each exhalation, the concentration of $N_2$ vs. time has a profile similar to that in Fig. 9. The concentration of $N_2$ coming out is low at the beginning of exhalation since that is the gas (mostly $O_2$) from the upper generations of the lung. The $N_2$ concentration goes up when $N_2$ from deep lung starts exiting. The slope of this final phase of the graph is not zero [84–87]. This is the slope of the "alveolar plateau." It has been found that this slope is indicative of asymmetry or heterogeneity in the lung. A 1D trumpet model, which effectively represents a symmetric lung with well-mixed gas, is known to give a near zero slope for the alveolar plateau unlike experimental data [86].

Models with some degree of asymmetric branching are shown to capture the slope of the alveolar plateau [63, 85–87]. This can be understood as follows. During inhalation of pure $O_2$, the concentration of $N_2$ in the deeper generations is high and at the mouth it is zero. When exhalation begins, the concentration of $N_2$ is convected out to the mouth. We can convert the concentration vs. generation number $N$ graph during inhalation to the concentration vs time graph at the mouth during exhalation by the following a transformation. For simplicity consider the steady state concentration vs. $N$ graph as the limiting profile during inhalation. The transformation basically finds the time $t_e$ it will take for the gas at location $x_e$ to reach the mouth:

$$N_e = \frac{\ln\left[1 - \frac{(1-\alpha)\ln(2\alpha\beta)}{\alpha\ln\alpha}\left(\frac{Qt_e}{A_0 L_0}\right)\right]}{\ln(2\alpha\beta)}.$$

(74)

Using this transformation, the concentration vs $N$ graph for $N_2$ during inhalation is converted to concentration vs time for $N_2$ at the mouth during exhalation. Now imagine the whole lung to be composed of two asymmetric branches with different Peclet numbers. Model each branch as a 1D trumpet. The trumpet model solution is applicable for each branch. The $N_2$ concentration coming out of each branch is mixed at the junction during exhalation. Upon analytically adding the $N_2$ concentration exhaled from each branch, the slope of the alveolar plateau is recovered as seen in Fig. 9. This slope arises due to the superposition of the convective and diffusive portions of the concentration profiles in the two branches [85, 87]. Models with more levels of branching will allow resolution of the phenomenon with greater fidelity [63].

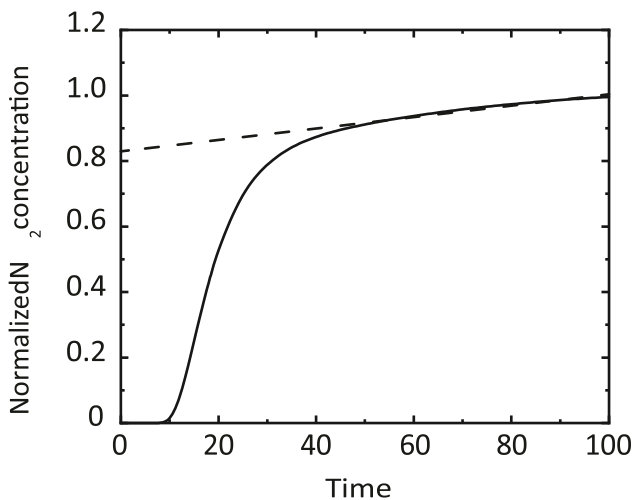

Figure 9: Nitrogen washout curve analytically calculated for two asymmetric branches merging. Dashed line represents the "phase III" slope of the washout curve.

### 5.6. Clinical relevance

The 1D models of droplet and particle transport have been utilised to computationally study pulmonary drug delivery and retention [6] with major focus on identifying physiological conditions conducive for delivering drugs to the deep lung. This is crucial for achieving systemic drug delivery. Drug delivery efficacy to the deep lung is observed to remain highest for $1 - 5\mu m$ aerosols. Drug deposition in the deep lung is observed to increase by a factor of 2, with respect to normal breathing, with doubling of the breathing time period. This suggests that breath control may be utilised as for enhancing drug delivery efficacy to the deep lung. Inhaled drug load reduces with increase in efficacy which can help in minimizing the side effects associated with drug inhalation.

The pathogen transport model, coupled with the droplet transport and immune response models, can be utilised to study the spatio-temporal progression of infections within the lung. Chakravarty et al. [36] utilised this model for understanding SARS-CoV-2 spread to the deep lung and proposed a new paradigm for this propagation called Reaerosolisation of Nasopharyngeal Mucosa (RNM). The model predicted that a severe infection (leading to pneumonia) develops within the deep lung within 2.5 - 7 days of symptom onset, in agreement with clinical findings. It was observed that immune responses, particularly antibodies and T-lymphocytes, play a critical role in preventing a severe infection. This also reinforces the role played by vaccination in managing a severe infection. The analysis also revealed that managing aerosolization of infected nasopharyngeal mucosa can be used as a potential strategy for minimizing infection spread and severity within the lung.

## 6. Outlook

**Improvement of 1D models and disease specific modeling.** One-dimensional models discussed in this perspective rely on reduced-order forms of various physical mechanisms. For example, models for deposition are used. Similarly, a better understanding of aerosolization within the airways will help develop aerosolization source terms in the 1D governing equations like there are deposition models used in it. Mucosal layer bridge formation within small airways ("airway closure") in the deep lung has been reported as a source of aerosolization within the lungs [88]. Shear layer atomization of the mucosal layer during coughing has been reproduced via simulations [32]. Mucosal shear flow during normal breathing is generally regarded as being stable. Yet, there has been evidence on increased exhaled aerosol during SARS-CoV-2 infection [4]. Shear flows of viscoelastic fluids that are linearly stable have been found to be nonlinearly unstable in the presence of large perturbations [89]. Nonlinear stability of the mucosal layer that is driven (and disturbed) by ciliary beating needs further investigation as a potential source of aerosolization.

The need for better aerosolization models is particularly important in the context of SARS-CoV-2 specific modeling since mucosal aerosolization might be significantly enhanced in SARS-CoV-2 patients [4, 8, 25, 32]. How does SARS-CoV-2 (and diseases in general) impact mucosal layer viscoelastic properties, and how does that impact the degree of aerosolization within the airways and in the nasopharyngeal region? How would that impact aerosol transport in the airways and would it increase the chances of deep lung infection as seen in SARS-CoV-2 patients?

Another well known airborne infection is pulmonary tuberculosis (TB) which continues to be a global menace. Sparse data are available regarding the mechanism underlying the occurrence of TB infection wherein it is intriguing as to why only 70% of individuals exposed do not get infected whereas only 30% develop infection [90, 91]. Careful experiments and fluid dynamics calculations would be important tools to gain these insights.

**Generalization to tree network.** Throughout this perspective, the 1D trumpet model is used to represent the entire lung. Different aspects, such as gas exchange, particle transport and deposition in the airways, mucus transport, particle transport in the mucus, and reaction kinetics for growth or mitigation of the infection are presented within the same trumpet model framework. As discussed in this perspective, several key insights into lung physiology have been reported through the years using this approach. While all aspects of lung physiology may not be captured by a 1D trumpet model for the entire lung (e.g. $N_2$ washout) there is scope for generalization. The 1D governing equations need not be used to represent the entire lung. Instead 1D equations could be used to represent branches or collection of branches in the lung tree network. Thus, modeling with various degrees of resolution could be done to gain more resolved insights into the physiology of a heterogeneous lung [63]. In this direction stochastic multi-path models of varying granularity has been the focus of one of our groups (Panchagnula) and have yielded significant insights.

**A virtual disease landscape.** In this perspective, the 1D governing equations were presented in terms of non-dimensional numbers. As far as the physics of lung function is concerned, the non-dimensional numbers represent the fundamental parameters or metrics on which lung physiology would depend. Consequently, normal or pathological patient groups should occupy different regions in this parameter space defining a virtual disease landscape (VDL). The existence of a VDL was shown for esophageal function in one of our groups (Patankar) [92] and should be explored in the future for the lungs. The non-dimensional numbers defining the VDL could be a systematic set of metrics or physiomarkers for disease classification and diagnosis as was shown for esophageal diseases [93]. This can help get mechanistic insights into pathologies. Disease progression could be tracked on the VDL, which coupled with machine learning techniques could be a powerful tool to predict future trajectory of diseases. Finally, the changes in physics-based parameters could be associated with biochemical alteration in the lungs. This connection between functional parameters space to the biochemical space can provide deeper insights into pathogenesis.

**Hybrid diagnostics tools.** Measuring the non-dimensional numbers on a patient specific basis to define the "mechanical health" of the lungs in the VDL remains a biomechanics challenge in clinical diagnostics. Reduced-order 1D models coupled with physiologic data from diagnostic tools and machine learning techniques could help measure the physiomarkers (e.g. non-dimensional numbers) for disease classification and diagnosis [92, 94, 95]. This technology could potentially help personalize patient care and treatment planning. Improved and suitably generalized 1D modeling could be an important tool in developing novel physics-machine-learning-clinical-data-based hybrid platforms that could be deployed at point-of-care.

## Appendix A

*1D analytical solution: exponential variation.* This section details the analytical approach for obtaining solution of Eq. 15 considering an exponential variation of the lung cross-sectional area. The total cross-sectional area ($A_x$) along the length ($x$) of the idealised lung geometry is approximated using an exponential function as

$$A_x = A_0 \exp(\beta x), \tag{A1}$$

where $A_0$ is the cross-sectional area at $x = 0$ and $\beta$ is the area-change factor. Integrating Eq. 15, we obtain -

$$\left( Q c_g - A_x D_g \frac{\partial c_g}{\partial x} \right) = K_1 \tag{A2}$$

where $K_1$ is the integration constant. Dividing the above equation with $A_x D_g$ and rearranging using Eq.A1, we obtain -

$$\frac{\partial c_g}{\partial x} - \frac{Q}{A_0 \exp(\beta x) D_g} c_g = - \frac{K_1}{A_0 \exp(\beta x) D_g}, \tag{A3}$$

which is of the form $y' + py = q$. The integrating factor for Eq. A3 is evaluated as

$$l = \exp \left[ \int p \, dx \right] = \exp \left[ - \frac{Q}{\beta A_0 D_g} \exp(-\beta x) \right] = \exp \left[ b \exp(-\beta x) \right], \tag{A4}$$

where $b = - \dfrac{Q}{\beta A_0 D_g}$. Utilising this, the integration is carried out as follows -

$$Z \, c_{g,x} l =$$
$$lq \, dx, \tag{A5}$$

$$\implies c_{g,x} \exp \left[ b \exp(-\beta x) \right] = \int \exp \left[ b \exp(-\beta x) \right] \left( - \frac{K_1}{A_0 \exp(\beta x) D_g} \right) dx = - \frac{K_1}{A_0 D_g} \int \exp \left[ b \exp(-\beta x) - \beta x \right] dx, \tag{A6}$$

$$\implies c_{g,x} \exp \left[ b \exp(-\beta x) \right] = \frac{K_1}{A_0 D_g} \frac{1}{\beta b} \exp \left[ b \exp(-\beta x) \right] + K_2, \tag{A7}$$

$$\implies c_{g,x} = \left[ \frac{K_1}{Q} \exp \left[ b \exp(-\beta x) \right] + K_2 \right] \exp \left[ - b \exp(-\beta x) \right]$$
$$= \frac{K_1}{Q} + K_2 \exp \left[ - b \exp(-\beta x) \right]. \tag{A8}$$

Equation A8 is solved subject to the following boundary conditions (see assumptions) -

$$c_{g,x}|_{x=0} = c_{g,0}, \tag{A9}$$

$$K_1|_{x=L} = A_L D_{ex}(c_{g,L} - c_{g,\infty}). \tag{A10}$$

Utilising the boundary conditions, we can write -

$$c_{g,0} = \frac{K_1}{Q} + K_2 \exp(-b) = \frac{K_1}{Q} + K_2 E_0, \tag{A11}$$

$$K_1 = \frac{A_L D_{ex} \left[ c_{g,0} \dfrac{E_L}{E_o} - c_{g,\infty} \right]}{\left[ 1 - \dfrac{A_L D_{ex}}{Q} \left( 1 - \dfrac{E_L}{E_0} \right) \right]}, \tag{A12}$$

where $E_0$ and $E_L$ represents the exponential terms in the above equations at $x = 0$ and $x = L$, respectively. Equation A12 is re-arranged using Eq. A1 to obtain the final expression of $K_1$ as

$$K_1 = \frac{A_L D_{ex}\left[c_{g,0} - c_{g,\infty}\dfrac{E_0}{E_L}\right]}{\left[\dfrac{E_0}{E_L} - \dfrac{A_L D_{ex}}{Q}\dfrac{E_0}{E_L} + \dfrac{A_L D_{ex}}{Q}\right]} = \frac{Q\left[c_{g,0} - c_{g,\infty}\dfrac{E_0}{E_L}\right]}{\left[\dfrac{Q}{A_0 D_{ex}}\dfrac{E_0}{E_L\exp(-\beta L)} - \dfrac{E_0}{E_L} + 1\right]}.$$ (A13)

$K_2$ is determined using Eq. A11 as

$$K_2 = \frac{c_{g,0} - \dfrac{K_1}{Q}}{E_0}.$$ (A14)

Substituting the expressions of $K_1$ and $K_2$ in Eq. A8 gives us the final solution in terms of $c_{g,x}$. This is expressed considering $c_{g,\infty} \to 0$ as follows

$$\lim_{c_{g,\infty},0}\frac{c_{g,x}}{c_{g,0}} = \phi_{g,x} = 1 + \frac{\left(\dfrac{E_0}{E_L} - \dfrac{E_x}{E_L}\right)\left[1 - \dfrac{Z}{\exp(-\beta L)}\right]}{1 - \dfrac{E_0}{E_L} + \dfrac{ZE_0}{E_L\exp(-\beta L)}}$$

$$= 1 + \frac{\left[\exp\left[Pe_{g,e}\left(1 - e^{-\beta L}\right)\right] - \exp\left[Pe_{g,e}\left(e^{-\beta x} - e^{-\beta L}\right)\right]\right]\left[1 - \dfrac{Z}{e^{-\beta L}}\right]}{1 - \exp\left[Pe_{g,e}\left(1 - e^{-\beta L}\right)\right]\left[1 - \dfrac{Z}{e^{-\beta L}}\right]}$$ (A15)

,

where $Pe_{g,e}\ \left(= \dfrac{QL_0}{A_0 D_g}\right)$ and $Z\ \left(= \dfrac{Q}{A_0 D_{ex}}\right)$ represents the Peclet number for gases (exponential variation) and the gas exchange parameter with the blood stream, respectively.

**Appendix B**

*1D analytical solution: power-law variation.* The length ($L_N$) and the total cross-sectional area ($A_N$) at each generation ($N$) of the idealised lung geometry can be approximated using a power-law function (see Section 1: Fig. 2 and Eq. 1). The length-change ($\alpha$) and area-change ($\beta$) factors in Eq. 1 are selected such that the computed length and area at each generation matches Weibel's morphometric data [12] as closely as possible (see Fig. 2a). The airway length ($x$), in terms of the lung generation number $N$, is given by

$$x_N = \frac{L_0(1 - \alpha^{N+1})}{1 - \alpha}.$$
(A16)

Since the length and area variation is assumed to be a function of $N$, the steady-state gas transport equation (Eq. 15) is re-written in terms $N$ as

$$\frac{\partial}{\partial N}\left( Qc_{g,N} - A_N D_g H_N \frac{\partial c_{g,N}}{\partial N} \right) = 0,$$
(A17)

using $H_N = \dfrac{\partial N}{\partial x_N}$. It is to be noted that although $N$ is an integer, it is treated as a continuous variable in all transport equations for computational convenience. Integrating, we obtain

$$\left( Qc_{g,N} - A_N D_g H_N \frac{\partial c_{g,N}}{\partial N} \right) = K_1,$$
(A18)

where $K_1$ is the integration constant. Dividing the above equation with $A_N D_g H_N$ and rearranging using Eq.1, we obtain

$$\frac{\partial c_{g,N}}{\partial N} - \frac{Q}{A_0(2\beta)^N D_g H_N} c_{g,N} = -\frac{K_1}{A_0(2\beta)^N D_g H_N},$$
(A19)

which is of the form $y^0 + py = q$ similar to that obtained for exponential variation. The solution technique used is also similar to that used for exponential variation. The integrating factor is evaluated as

$$l = \exp\left[\int p dN\right] = \exp\left[\int -\frac{Q}{A_0(2\beta)^N D_g H_N} dN\right] = \exp\left[-\frac{Q}{A_0(2\beta)^N D_g H'} \frac{(\alpha/2\beta)^N}{\ln(\alpha/2\beta)}\right] = \exp[ba^N], \quad \text{(A20)}$$

where $H' = \dfrac{1 - \alpha}{-L_0 \alpha \ln(\alpha)\alpha^N}$, $a = \alpha/2\beta$ and $b = -\dfrac{Q}{A_0(2\beta)^N D_g H' \ln(\alpha/2\beta)}$. Utilising this, the integration is carried out as follows -

$$c_{g,N}l = \int lq dN,$$
(A21)

$$\implies c_{g,N}\exp\left[ba^N\right] = \int \exp\left[ba^N\right]\left( -\frac{K_1}{A_0(2\beta)^N D_g H_N} \right) dN = -\frac{K_1}{A_0 D_g H'} \int a^N \exp\left[ba^N\right] dN, \quad \text{(A22)}$$

$$\implies c_{g,N}\exp\left[ba^N\right] = -\frac{K_1}{A_0 D_g H'} \frac{\exp\left[ba^N\right]}{b\ln(a)} + K_2,$$
(A23)

$$\implies c_{g,N} = \left[ -\frac{K_1}{A_0 D_g H'} \frac{\exp\left[ba^N\right]}{b\ln(a)} + K_2 \right]\exp\left[ -ba^N \right],$$

$$= \left[ -\frac{K_1}{A_0 D_g H'} \frac{A_0 D_g H'\ln(\alpha/2\beta)}{-Q\ln(\alpha/2\beta)} \right] + K_2\exp\left[ -ba^N \right]$$

$$= \frac{K_1}{Q} + K_2\exp\left[ -\frac{QL_0\alpha\ln(\alpha)}{A_0 D_g(1 - \alpha)\ln(\alpha/2\beta)} \left( \frac{\alpha}{2\beta} \right)^N \right].$$
(A24)

Equation A24 is solved subject to the boundary conditions discussed before (see assumptions). Mathematically, these boundary conditions are expressed in terms of $N$ as -

$$c_{g,N}|_{N=0} = c_{g,0},$$ (A25)

$$K_1|_{N=M} = A_M D_{ex}(c_{g,M} - c_{g,\infty}),$$ (A26)

where $M = 23$ i.e. the terminal lung generation. Utilising the boundary conditions, we can write

(A27) $$c_{g,0} = \frac{K_1}{Q} + K_2\exp\left[-\frac{QL_0\alpha\ln(\alpha)}{A_0 D_g(1-\alpha)\ln(\alpha/2\beta)}\left(\frac{\alpha}{2\beta}\right)^0\right] = \frac{K_1}{Q} + K_2 E_0 \quad ,$$

$$\implies K_1 = \frac{A_M D_{ex}\left[c_{g,0}\dfrac{E_M}{E_o} - c_{g,\infty}\right]}{\left[1 - \dfrac{A_M D_{ex}}{Q}\left(1 - \dfrac{E_M}{E_0}\right)\right]},$$ (A28)

where $E_M$ and $E_0$ represents the exponential term in Eq. A27 for at $N = M$ and $N = 0$, respectively. Equation A28 is re-arranged using Eq. 1 to obtain the final expression of $K_1$ as

$$K_1 = \frac{A_M D_{ex}\left[c_{g,0}\dfrac{E_M}{E_o} - c_{\text{inf}}\right]}{\left[1 - \dfrac{A_M D_{ex}}{Q}\left(1 - \dfrac{E_M}{E_0}\right)\right]}.$$ (A29)

We define the following

$$\delta = \frac{\alpha}{2\beta},\ Pe_{g,pl} = \frac{QL_0\alpha\ln(\alpha)}{A_0 D_g(1-\alpha)\ln(\alpha/2\beta)},\ Z = \frac{Q}{A_0 D_{ex}},$$ (A30)

such that we can write $E_N = \exp(-Pe_{g,pl}\delta^N)$. Eq. A29 can, thus, be written as

$$K_1 = \frac{A_M D_{ex}\left[c_{g,0}\dfrac{\exp(-Pe_{g,pl}\delta^M)}{\exp(-Pe_{g,pl}\delta^0)} - c_{g,\infty}\right]}{\left[1 - \dfrac{A_M D_{ex}}{Q}\left(1 - \dfrac{\exp(-Pe_{g,pl}\delta^M)}{\exp(-Pe_{g,pl}\delta^0)}\right)\right]},$$ (A31)

which can further be re-arranged using Eq. 1 as

$$K_1 = \frac{Q\left[c_{g,0} - c_{g,\infty}\exp\left(Pe_{g,pl}(\delta^M - 1)\right)\right]}{\left[1 - \exp\left(Pe_{g,pl}(\delta^M - 1)\right) + \dfrac{Q}{A_0 D_{ex}}\dfrac{\exp\left(Pe_{g,pl}(\delta^M - 1)\right)}{(2\beta)^M}\right]}.$$ (A32)

Using Eq. A27, $K_2$ can be evaluated as

$$K_2 = c_{g,0}\exp(Pe_{g,pl}) - \frac{K_1}{Q}\exp(Pe_{g,pl}),$$ (A33)

Substituting the magnitudes of $K_1$ (Eq. A32) and $K_2$ (Eq. A32) in Eq. A24 provides us with the following expression

$$\frac{c_{g,N}}{c_{g,0}} = \frac{\left[1 - \frac{c_{g,\infty}}{c_{g,0}} \exp\left(Pe_{g,pl}(\delta^M - 1)\right)\right]\left[1 - \exp\left(-Pe_{g,pl}(\delta^N - 1)\right)\right]}{\left[1 - \exp\left(Pe_{g,pl}(\delta^M - 1)\right) + Z\frac{\exp\left(Pe_{g,pl}(\delta^M - 1)\right)}{(2\beta)^M}\right]} + \exp\left(-Pe_{g,pl}(\delta^N - 1)\right), \quad (A34)$$

which can be re-arranged considering $c_{g,\infty} \to 0$ to give the final form of Eq. A24 as

$$\lim_{c_{g,\infty} \to 0} \frac{c_{g,N}}{c_{g,0}} = \phi_{g,N} = 1 + \frac{\left[\exp\left(Pe_{g,pl}(\delta^N - 1)\right) - \exp\left(Pe_{g,pl}(\delta^M - \delta^N)\right)\right]\left[1 - \frac{Z}{2\beta^M}\right]}{1 - \exp\left(Pe_{g,pl}(\delta^M - 1)\right) + Z\frac{\exp\left(Pe_{g,pl}(\delta^M - 1)\right)}{(2\beta)^M}}. \quad (A35)$$

**Appendix C**

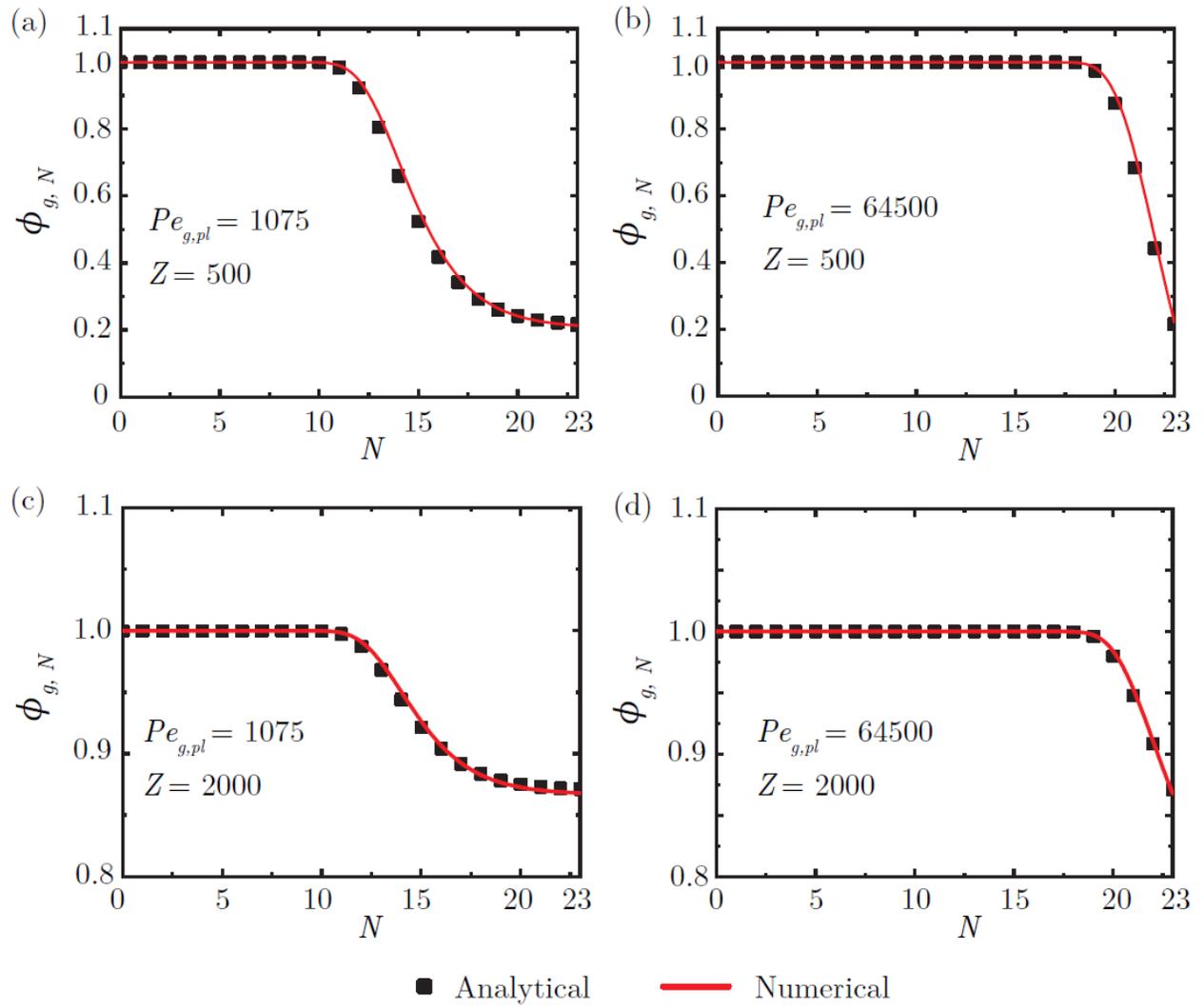

Figure A1: Comparison of numerical prediction of steady-state $\varphi_{g,N}$ with the analytical prediction (using the power-law model) for different combinations of $Pe_{g,pl}$ and $Z$ covering the entire range of the parameters considered in this analysis. The results are shown considering $St_\alpha = 0.01$, $\alpha = 0.73, \beta = 0.71$. Note the change in scales between the top row and bottom row figures.

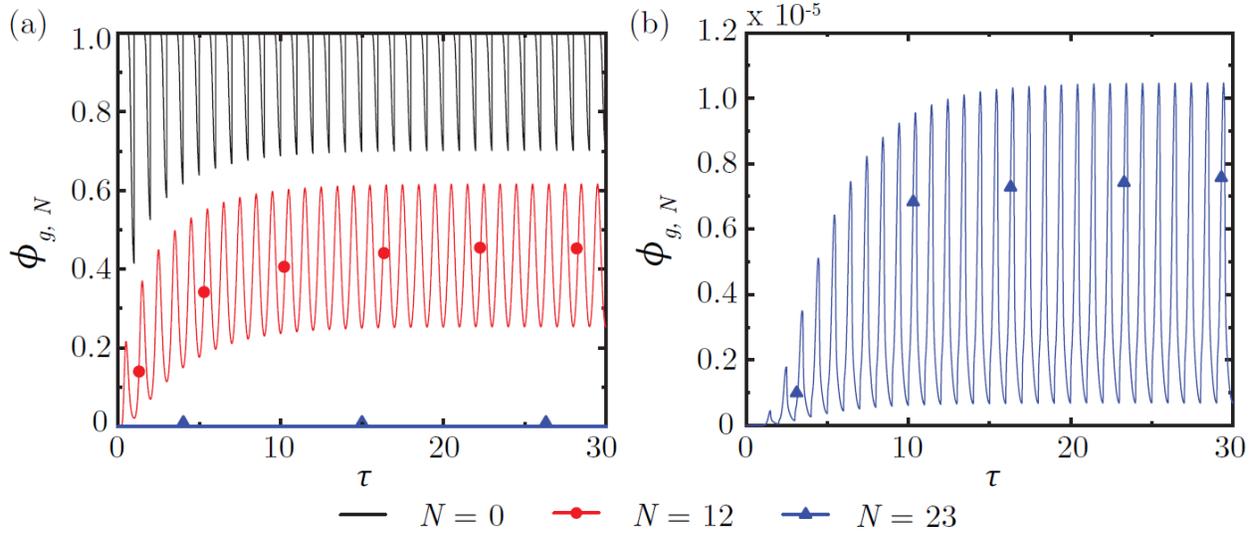

Figure A2: Temporal change in $\varphi_{g,N}$ at different lung generations for a physiologically realistic cyclic flow situation and considering a continuous gas source at the entrance to trachea. The results are shown for $Pe_g = 64500$, $Z = 500$, $St_a = 0.01$, $\alpha = 0.73$, $\beta = 0.71$

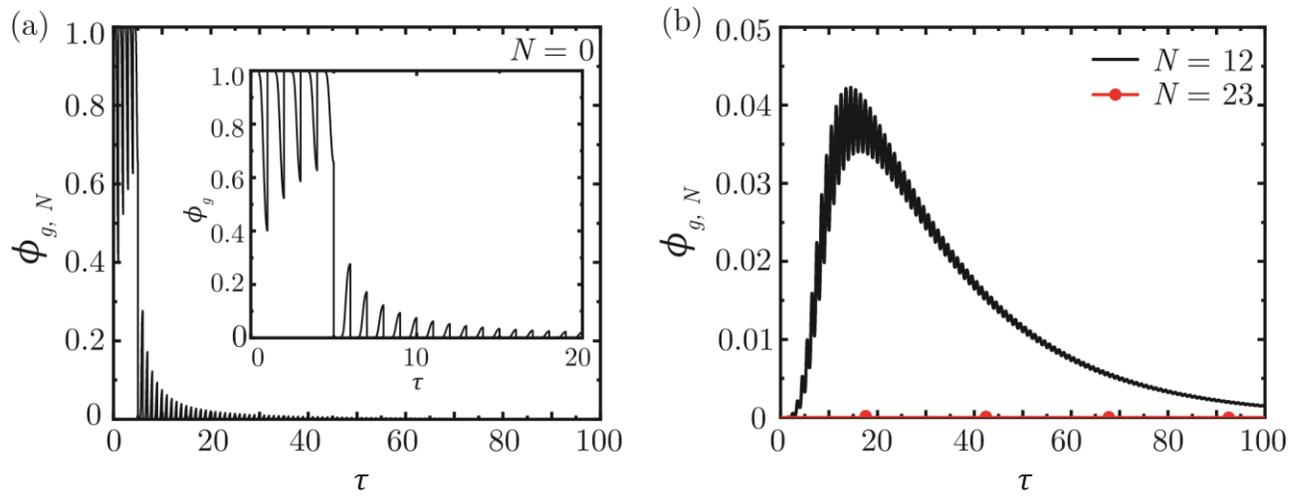

Figure A3: Temporal change in $\varphi_{g,N}$ at different lung generations for a physiologically realistic cyclic flow situation and considering finite availability of gas source at the entrance to trachea. The results are shown for $Pe_g = 64500$, $Z = 500$, $St_a = 0.01$, $\alpha = 0.73$, $\beta = 0.71$ and $\tau_{exp} = 5$.

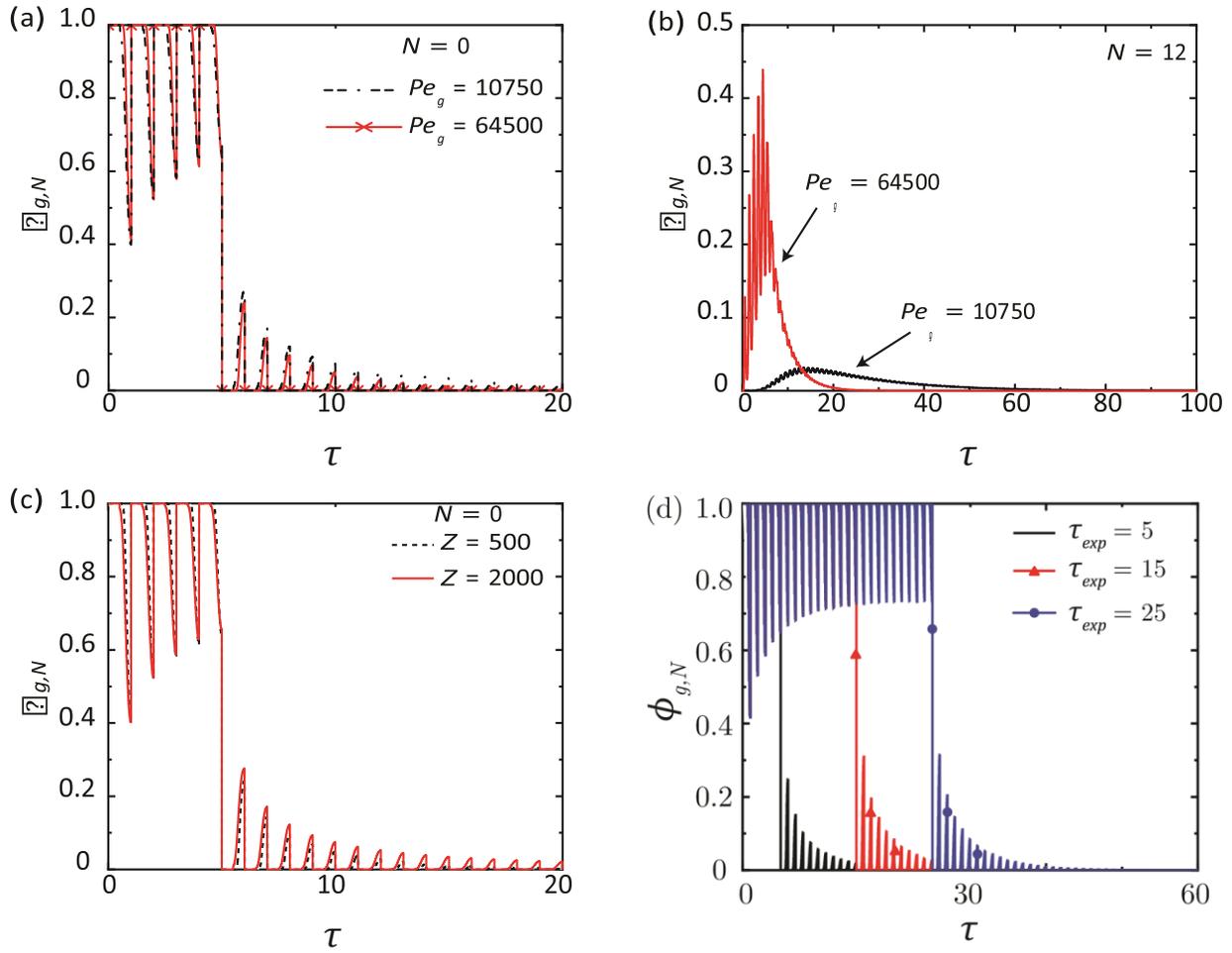

Figure A4: Temporal change in $\varphi_{g,N}$ for two different $Pe_g$ considering $Z = 500$, $St_a = 0.0095$ and $\tau_{exp} = 5$ at (a) $N = 0$ and (b) $N = 12$. (c) Temporal change in $\varphi_{g,N}$ for two different $Z$ considering $Pe_g = 64500$, $St_a = 0.0095$ and $\tau_{exp} = 5$ at $N = 0$ (d) Temporal change in $\varphi_{g,N}$ for three different $\tau_{exp}$ considering $Pe_g = 64500$, $St_a = 0.0095$ and $Z = 500$ at $N = 0$. The results are shown for $\alpha = 0.73, \beta = 0.71$.

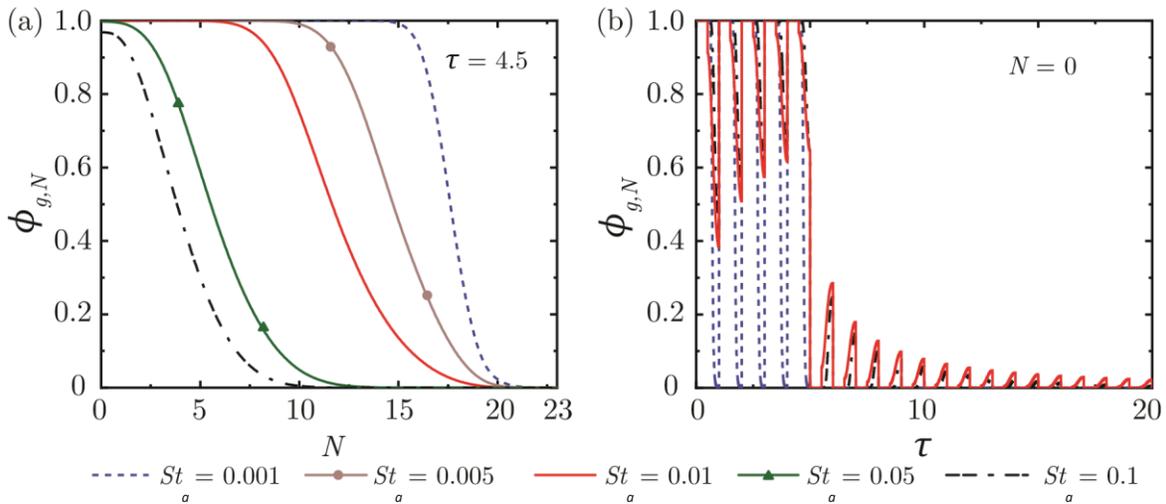

Figure A5: (a) Progression of $\varphi_{g,N}$ into the lung for various $St_a$ at $\tau = 4.5$ and (b) Temporal change in $\varphi_{g,N}$ at $N = 0$ for various $St_a$. The results are shown for $Pe_g = 64500$, $Z = 2000$, $\tau_{exp} = 5$, $\alpha = 0.73, \beta = 0.71$.

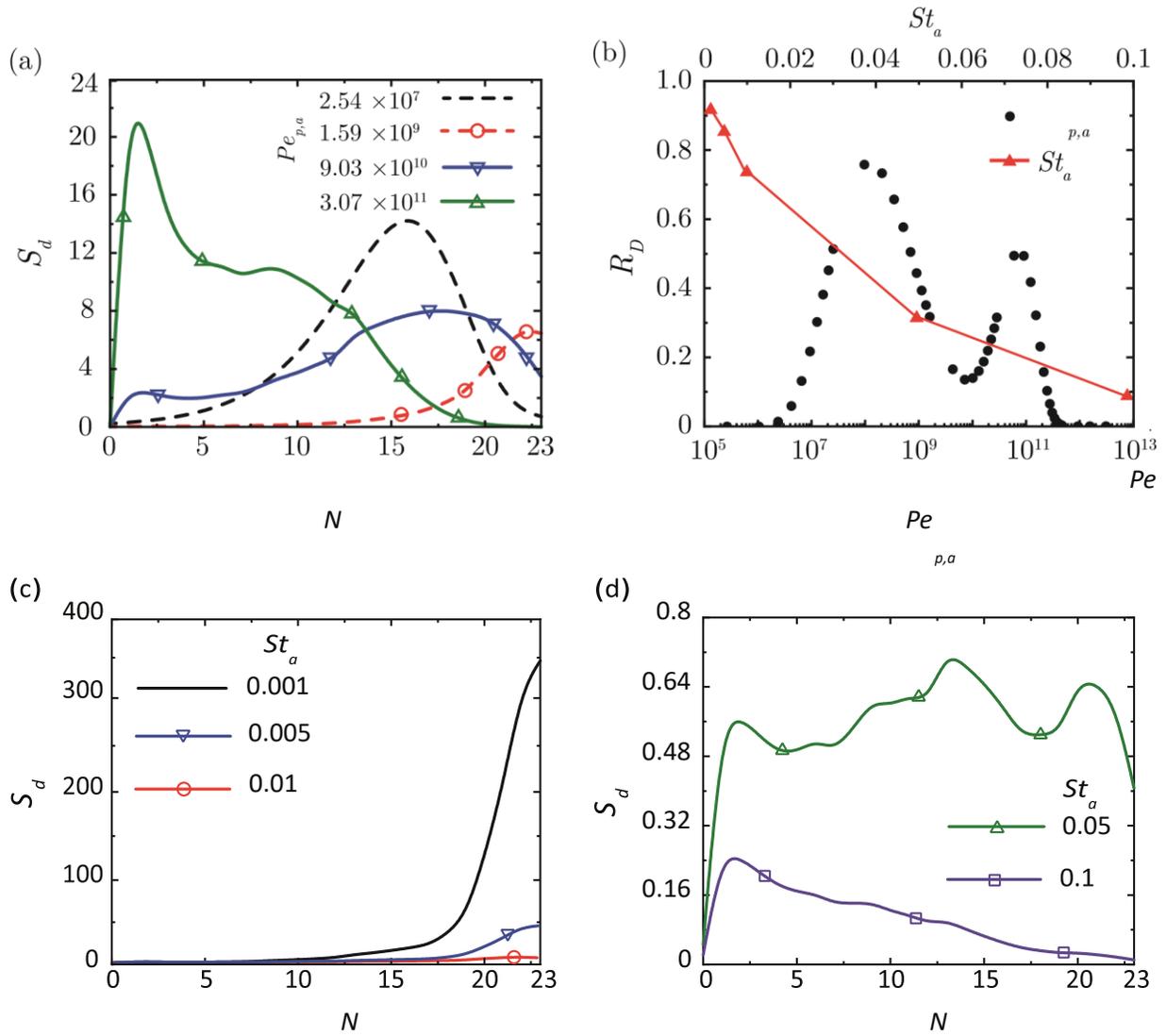

Figure A6: (a) Total particle deposition ($S_d$) within the lung for different $Pe_{p,a}$ considering $St_a = 0.0095$ and $\tau_{exp} = 5$. (b) Ratio ($R_D$) of particles deposited in the deep lung ($N > 17$) to that deposited in the whole lung with change in $Pe_{p,a}$ and $St_a$ (c-d) Total particle deposition ($S_d$) within the lung for different $St_a$ considering $Pe_{p,a} = 2.85 \times 10^{10}$ and $\tau_{exp} = 5$. Figure reproduced with permission from Chakravarty et al. [6].

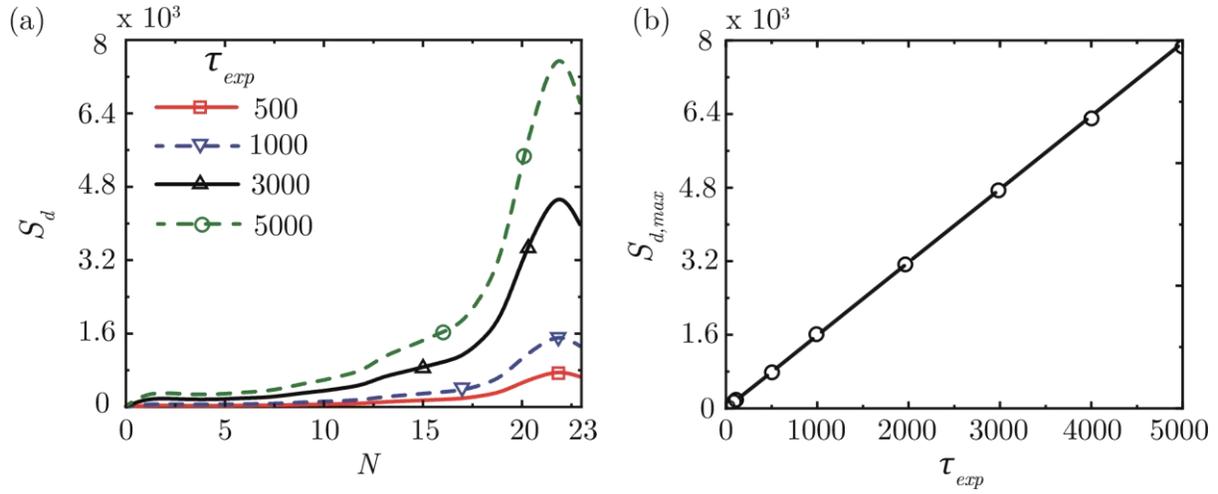

Figure A7: (a) Total deposition ($S_d$) of particles within the lung at the end of exposure for different exposure duration ($\tau_{exp}$) and (b) Increase in maximum particle deposition ($S_{d,max}$) with rise in exposure time. The results are shown for $Pe_{p,a} = 2.85 \times 10^{10}$ and $St_a = 0.0095$. Figure reproduced with permission from Chakravarty et al. [6].

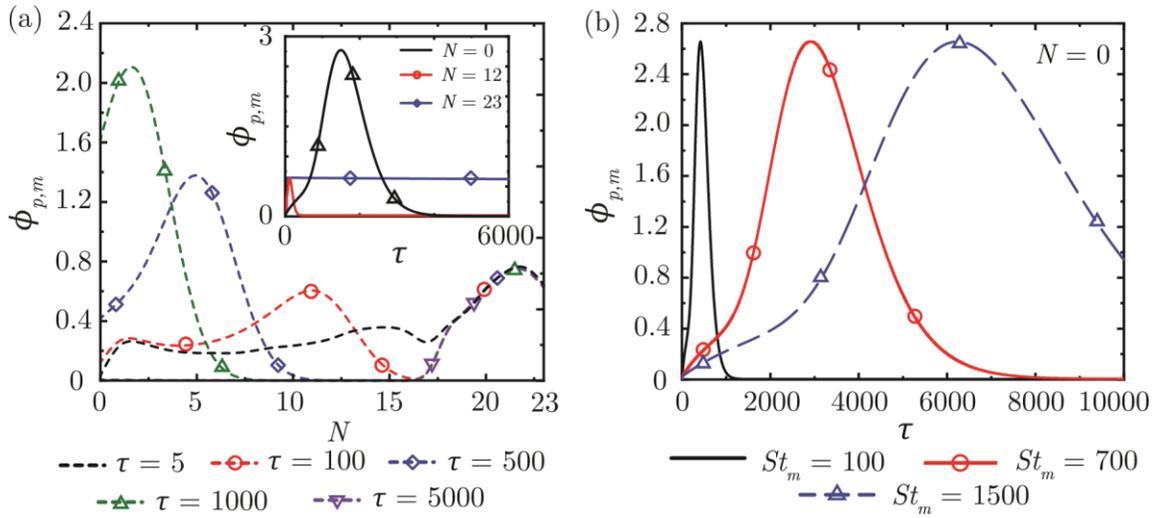

Figure A8: (a) $\varphi_{p,m}$ within the lung at different time instances and (inset) variation of $\varphi_{p,m}$ with time at three different lung generations ($N = 0,12,23$) considering $Pe_{p,a} = 2.85 \times 10^{10}$, $St_a = 0.0095$, $Pe_{p,m} = 4.56 \times 10^7$, $St_m = 359.7122$, $\tau_{exp} = 5$. (b) Temporal change in $\varphi_{p,m}$ at $N = 0$ for different $St_m$ other parameters remaining same. Figure reproduced with permission from Chakravarty et al. [6].

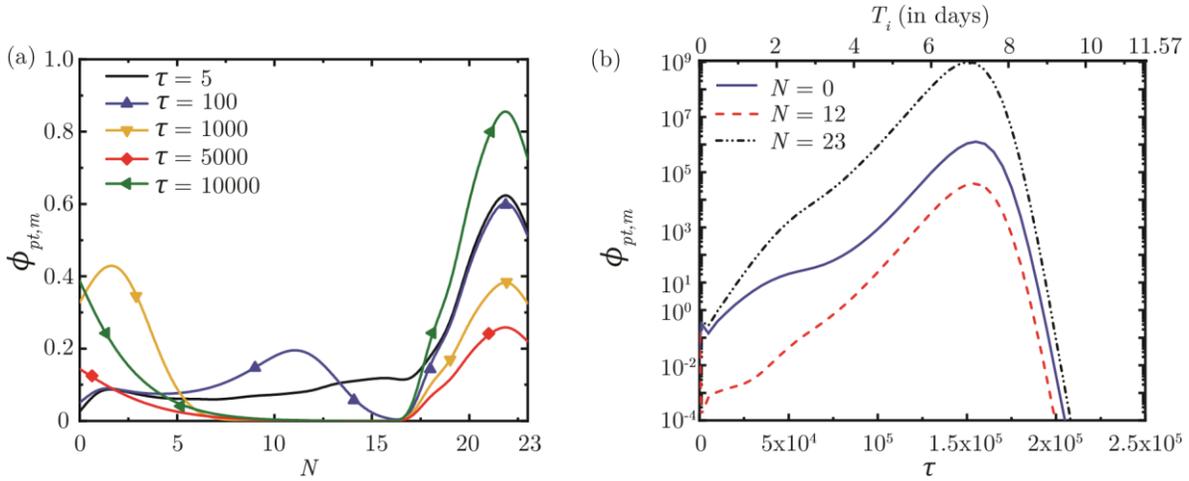

Figure A9: (a) Dimensionless pathogen concentration in mucus ($\varphi_{pt,m}$) within the LRT at different dimensionless time instances post infection onset and (b)Temporal change in $\varphi_{pt,m}$ at different spatial locations within the LRT ($N$ is the lung generation). The results are shown for a SARS-CoV-2 infection. The results in (b) are also shown with respect to a dimensional time ($T_i$) post infection onset considering breathing time period ($T_b$ = 4s). Figure reproduced with permission from Chakravarty et al. [36].

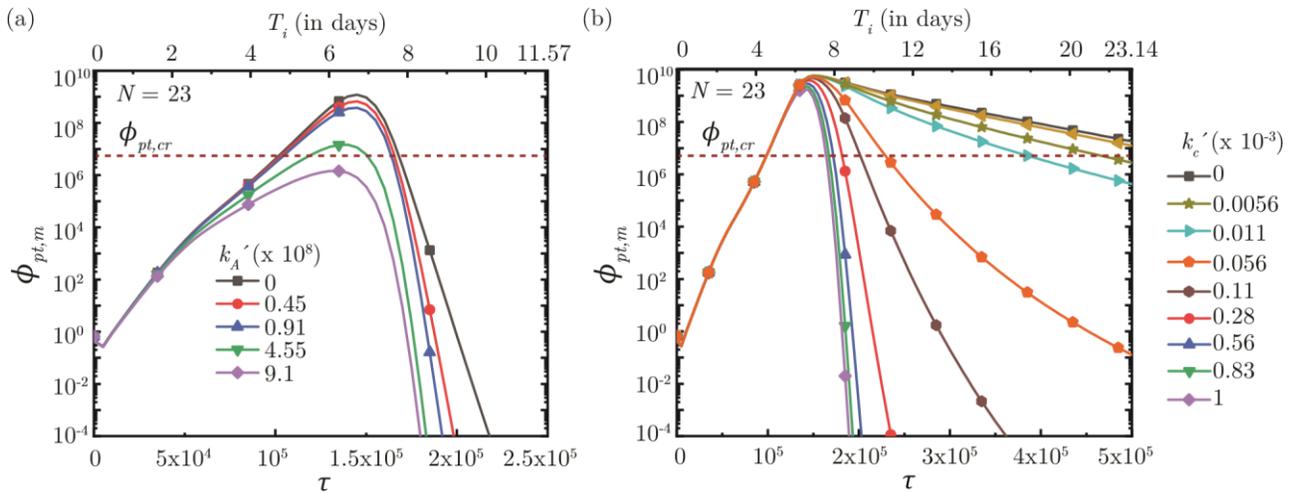

Figure A10: Temporal change in pathogen concentration ($\varphi_{pt,m}$) at $N$ = 23 (deep lung) with change in (a) binding affinity of antibodies ($k_A{}^0$) and (b) the rate at which cytotoxic T-lymphocytes eliminate the infected cells ($k_c{}^0$). The results are shown for SARS-CoV-2 infection with respect to dimensionless time ($\tau$) as well as a dimensional time ($T_i$) post infection onset with the breathing time period ($T_b$ = 4s). The dotted line indicates the critical SARS-CoV-2 load in the deep lung required for pneumonia onset. Figure reproduced with permission from Chakravarty et al. [36].

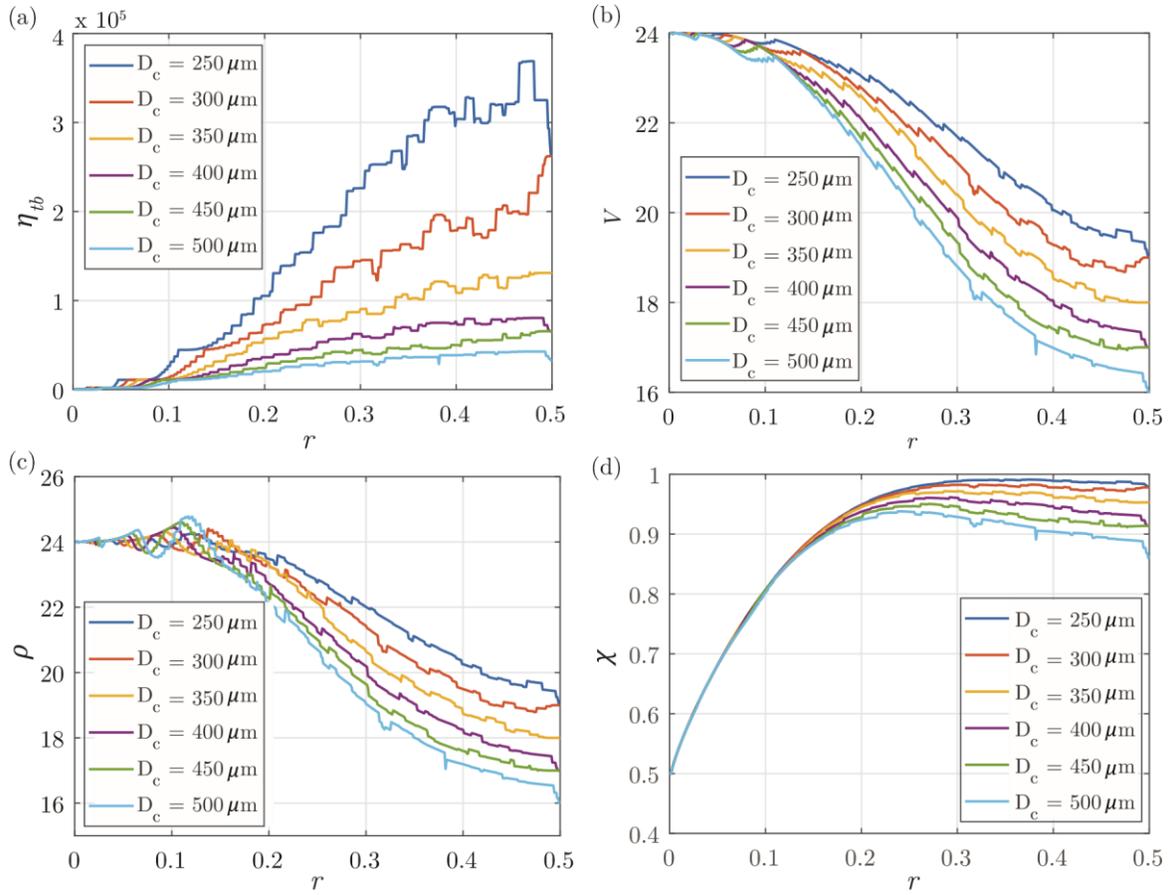

Figure A11: Variation of (a) number of terminal bronchioles ($\eta_{tb}$), (b) volume occupied by the bronchial tree ($V$), (c) resistance offered to fluid flow ($\rho$) and (d) particle filtration efficiency ($\chi$) as function of degree of asymmetry ($r$) for different values of cut-off diameter ($D_c$). Reproduced with permission from Kundu et al. [75].